\documentclass[prb,twocolumn,superscriptaddress,float,aps]{revtex4-2}
\usepackage{bm,color,amsmath,amssymb,mathrsfs,latexsym,graphicx,psfrag,tabularx}

\newcommand{\beq}{\begin{equation}}
\newcommand{\eneq}{\end{equation}}

\begin{document}

\title{Electrically Controlled Anomalous Hall Effect and Orbital Magnetization in Topological Magnet MnBi$_2$Te$_4$}
\author{Ruobing Mei}
\affiliation{Department of Physics, The Pennsylvania State University, University Park,  Pennsylvania 16802, USA}
\author{Yi-Fan Zhao}
\affiliation{Department of Physics, The Pennsylvania State University, University Park,  Pennsylvania 16802, USA}
\author{Chong Wang}
\affiliation{Department of Materials Science and Engineering, University of Washington, Seattle, WA 98195, USA}
%\affiliation{Department of Physics, University of Washington, Seattle, Washington, 98195, USA}
\author{Yafei Ren}
\affiliation{Department of Materials Science and Engineering, University of Washington, Seattle, WA 98195, USA}
%\affiliation{Department of Physics, University of Washington, Seattle, Washington, 98195, USA}
\author{Di Xiao}
\affiliation{Department of Materials Science and Engineering, University of Washington, Seattle, WA 98195, USA}
\affiliation{Department of Physics, University of Washington, Seattle, Washington, 98195, USA}
\author{Cui-Zu Chang}
\affiliation{Department of Physics, The Pennsylvania State University, University Park,  Pennsylvania 16802, USA}
\author{Chao-Xing Liu}
\affiliation{Department of Physics, The Pennsylvania State University, University Park,  Pennsylvania 16802, USA}
% \email{cxl56@psu.edu}

\begin{abstract}
We propose an intrinsic mechanism to understand the even-odd effect, namely opposite signs of anomalous Hall resistance and different shapes of hysteresis loops for even and odd septuple layers (SLs), of MBE-grown MnBi$_2$Te$_4$ thin films with electron doping. The non-zero hysteresis loops in the anomalous Hall effect and magnetic circular dichroism for even-SLs MnBi$_2$Te$_4$ films originate from two different anti-ferromagnetic (AFM) configurations with 
%opposite magnetoelectric coefficients that give rise to 
different zeroth Landau level energies of surface states. The complex form of the anomalous Hall hysteresis loop can be understood from two magnetic transitions, a transition between two AFM states followed by a second transition to the ferromagnetic state. 
Our model also clarifies the relationship and distinction between axion parameter and magnetoelectric coefficient, and shows an even-odd oscillation behavior of magnetoelectric coefficients in MnBi$_2$Te$_4$ films. 
\end{abstract}
  
\maketitle

%\section*{Introduction}

{\it Introduction - }The recent discovery of MnBi$_2$Te$_4$ (MBT) \cite{otrokov2017highly,yan2019crystal,gong2019experimental,otrokov2019prediction,wu2019natural,shi2019magnetic,he2020mbt,liu2021magnetic}
%, a tetradymite-type anti-ferromagnetic compound, 
provides an excellent platform to explore the interplay between topological physics and magnetism\cite{chang2020marriage,chang2023colloquium}. Exotic magnetic topological phases, including the quantum anomalous Hall (QAH) state \cite{otrokov2019unique,li2019intrisic}, axion insulator (AI) \cite{otrokov2019unique,li2019intrisic,zhang2019topological} and  M\"obius insulator \cite{zhang2020mobius}, have been theoretically predicted. For bulk materials, the A-type anti-ferromagnetism, namely ferromagnetic coupling in one septuple layer (SL) and anti-ferromagnetic (AFM) coupling between two adjacent SLs, has been unambiguously established through magnetic susceptibility \cite{gong2019experimental,otrokov2019prediction} and neutron diffraction experiments \cite{yan2019crystal}. 
Topological Dirac surface states have also been observed in angular-resolved photon emission spectroscopy \cite{gong2019experimental,chen2019topological,estyunin2020signatures,nevola2020coexistence,yan2021origins}, although the existence of magnetic gap is still under debate \cite{otrokov2019prediction,rienks2019large, chen2019topological,swatek2020gapless}. These experiments confirmed the coexistence of magnetic order and topological band structure in bulk MBT. 

The situation of MBT thin films however is subtle. Theoretically, an even-odd effect was predicted for {\it insulating} MBT films \cite{otrokov2019unique,li2019intrisic,zhang2019topological}. The QAH state can exist for odd SLs 
%with the parallel magnetization on the top and bottom SLs 
while the AI state 
\cite{qi2011topological,xiao2018realization,xu2019higher,chen2021using,gu2021spectral,liu2020robust,mogi2017magnetic,mogi2017tailoring,li2021critical,song2021delocalization,fijalkowski2021any,pierantozzi2022evidence,yu2019magnetic,yue2019symmetry,sekine2021axion,li2010dynamical,chen2023side,otrokov2019unique,li2019intrisic} appears
 for even SLs. 
 %with the anti-parallel magnetization. 
%  of the top and bottom SLs are anti-parallel, leading to a zero net magnetization and the axion insulator state
% in which the resistance shows a zero Hall plateau \cite{}. 
Later experiments combining reflective magnetic circular dichroism (RMCD) and anomalous Hall (AH) measurements, however, challenged this scenario \cite{chen2019intrinsic,deng2020quantum,ovchinnikov2021intertwined,yang2021odd,cai2022electric,qiu2023axion}. 
%through  in exfoliated MBT flakes . 
The corresponding hysteresis loops are not synchronized. Specifically, RMCD signals exhibit a clear hysteresis loop for odd SLs whereas the AH hysteresis loop is almost invisible. For even SLs, a small zero-field RMCD signal was reported, whereas a clear AH hysteresis loop was found. These experimental findings indicate the complexity of real materials where the chemical and magnetic environments that depend on individual sample qualities are important.
%Thus, it is desirable to look for transport signatures that are intrinsically correlated to the AFM order and can be repeatably observed in experiments. 

\begin{figure}
\includegraphics[width=\columnwidth]{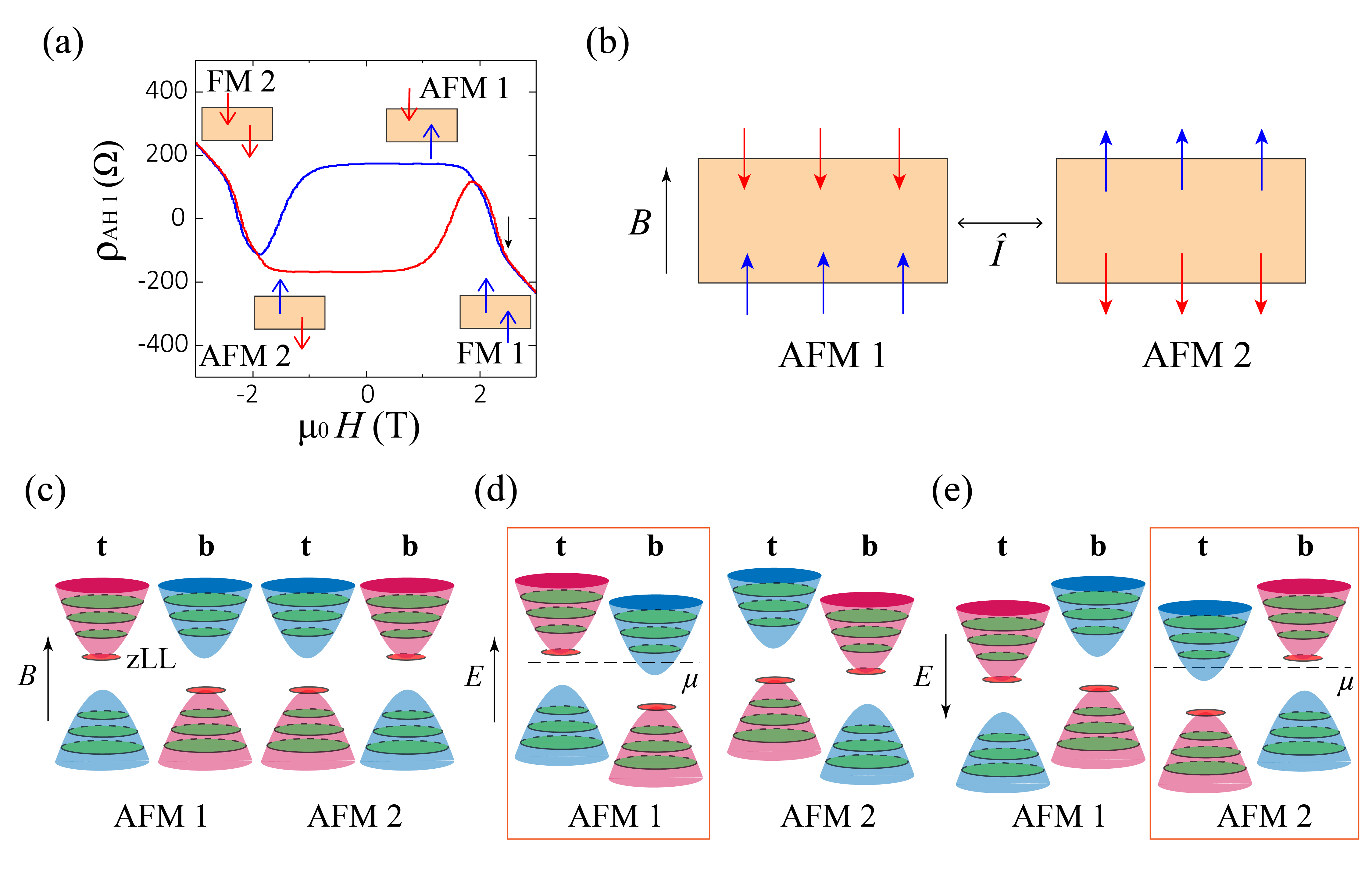}
\caption{(a) Experimental measurement of AH resistance $\rho_{yx}$ as a function of magnetic field $\mu_0 H$ in a 2 SL MBT film. The spin-flop field is around 2.3 T, indicated by the black arrow. 
%Here only the contribution to the Hall resistance from the intrinsic AH effect in MBT is included while other contributions have been excluded. %, e.g. the AH contributions from Mn doped (Bi,Sb)$_2$Te$_3$ and ordinary Hall effect, have been excluded. % The intrinsic AH contribution for other even SL MBT films is similar to that of 2 SL MBT films. 
See 
%\emph{Zhao et al.} 
Ref. \cite{zhao2021even} and Appendix Sec.I \cite{sm2023} for more details. The favored AFM states denoted here are for the case with $V_0>0$. When $V_0<0$, AFM1 and AFM2 are switched. (b) Magnetization configurations of even SL MBT films. % Under an external magnetic field, there are two possible AFM configurations related by inversion symmetry, which are energetically equivalent. 
(c) Illustration of the two-surface-state model for $g>0$ and $B>0$, where ``t" and ``b" stand for top and bottom surfaces, respectively. Each band is labeled with blue or red color, which represents positive or negative AH sign. The zeroth Landau levels are shown in red and the other Landau levels are shown in green. (d-e) Illustration of two surface states with an electric field (d) $E>0$ and (e) $E<0$. %(d) When an electric field $E$ is present, the top and bottom surfaces split in energies. For a negative $E$ ($V_0>0$), the occupied zLL in AFM1 is lower in energy than that in AFM2, and therefore, AFM1 becomes the favored configuration. At electron doping, the conduction band of the bottom surface contributes positive AH sign (blue color). (e) For a positive $E$ ($V_0<0$), AFM2 is the favored state also with positive AH sign at electron doping. 
}
\label{fig:fig1}
\end{figure}

Recently, another type of even-odd effect was found in {\it metallic} MBT films grown by molecular beam epitaxy (MBE)
%in the electron doping regime 
\cite{zhao2021even}. Although the metallic samples with both even and odd SLs show AH hysteresis loops, the loop shapes are clearly distinct. The AH hysteresis loops can be decomposed into two AH components. One behaves the same for even and odd SLs, coming from minor Mn-doped Bi$_2$Te$_3$. The other is from dominant MBT phase, and (1) the signs of zero-field AH resistance are opposite for even and odd SLs; (2) for even SLs, AH sign reverses twice around spin-flop transition between AFM and canted AFM states in Fig. \ref{fig:fig1}(a), reproducing the measurements in %\emph{Zhao et al.} 
Ref. \cite{zhao2021even}, while no such behavior occurs for odd SLs (Appendix Sec.I \cite{sm2023}). Similar transport data was also shown in even-SL MBT fabricated by mechanical exfoliation \cite{chen2019intrinsic,ovchinnikov2021intertwined,gao2021layer,zhang2020experimental}. 
% In \emph{Chen et al.} \cite{Chen2019}, the opposite AH signs for even and odd SLs were attributed to the competition between intrinsic 
% %(Berry curvature contribution) 
% and extrinsic 
% %(impurity scattering contribution) 
% AH effects \cite{ZhangS2020}. However, 
Given different disorder levels for the samples prepared with different methods, we here explore intrinsic mechanism for this even-odd effect. 

In this work, we provide a theoretical understanding of AH hysteresis loops in MBT films based on a two-surface-state model and a four-band thin film model. Our theory suggests that the transition between two nearly degenerate AFM states (Fig. \ref{fig:fig1}(b)) can provide a consistent understanding of both the opposite signs between even and odd SLs and the complex AH hysteresis loop of even SLs in Fig. \ref{fig:fig1}(a). Furthermore, we interpret our results with orbital magnetization \cite{huang2004spin,lu2019superconductors,sharpe2019emergent,he2020giant} and extract the magnetoelectric coefficient, which approaches the bulk value determined by the axion parameter as the layer thickness increases for even SLs and stays zero for odd SLs. We clarify the relationship and distinction between axion parameter and magnetoelectric coefficient.

{\it Anomalous Hall hysteresis loop - }
For even SL MBT, there are two degenerate AFM configurations, labelled as AFM1 and AFM2 (Fig. \ref{fig:fig1}(b)), related by either inversion $\hat{I}$ or time reversal symmetry $\hat{T}$. This degeneracy is lifted in the presence of both magnetic and electric fields from asymmetric substrates or electric gates \cite{gao2021layer,cai2022electric}.
%To make the above argument more concrete, let us 
To describe AFM configurations, we consider a two-surface-state model with the Hamiltonian $H=H_M+H_e+H_{e-M}$,
% \beq H=H_M+H_e+H_{e-M},
% \eneq
where $H_M$ describes the magnetization part, $H_e$ is for the electron part and $H_{e-M}$ gives the coupling between electrons and magnetization \cite{li2022identifying,mills1968surface,ovchinnikov2021intertwined}. 
The explicit forms of $H_M$, $H_e$, $H_{e-M}$ are given in Appendix Sec.II.A \cite{sm2023}. %\addCXL{The discussion of $H_M$ should be included the appendix}. 
Magnetic simulations in 
%\emph{Zhang et al.} 
Ref. \cite{zhang2019topological} and %\emph{Ovchinnikov et al.} 
\cite{ovchinnikov2021intertwined} suggest that the ground state of $H_M$ is the out-of-plane AFM configurations, %with $\bm{m}_{s1}=(0,0,\pm1)$ and $\bm{m}_{s2}=-\bm{m}_{s1}$ 
namely AFM1 and AFM2 in Fig. \ref{fig:fig1}(b) that are degenerate under $H_M$, at low magnetic fields $B$. The FM state is energetically favored at larger $B$. 
%, with magnetization vectors aligned with $B$.
% Thus, the magnetization energy for these two AFM states is $E_M=-J-K$. The energy of AFM states is independent of magnetic field $B$ due to the zero total magnetization $\bm{m}_{s1}+\bm{m}_{s2}=0$. On the other hand, the FM state has the magnetization energy $E_M=J-K\pm 2M_sB$, where $\pm$ selects the FM state with magnetization vectors aligned with $B$ as the favored configuration. Thus, for $J>0$, the AFM states have lower energy while the FM state can be energetically favored at a large $B$. 

The ground state energy for AFM1 and AFM2 can be distinguished by including electron energy of $H_e+H_{e-M}$, which involves Landau level (LL) spectrum under $B$ (Appendix Sec.II.A \cite{sm2023}).
%For Dirac surface states, besides the normal LLs, given by $\varepsilon_{\mu,\nu}^N =\mu\sqrt{2v_f^2N/l_c^2+(g M_s)^2} +\nu V_0/2$, where $\mu,\nu=\pm1$, $N=1,2,\cdots$ representing the $N$th LL and $l_c=\sqrt{\frac{\hbar}{eB}}$ is the magnetic length, there are additional zeroth LLs (zLLs), given by $\varepsilon_{1,\lambda}^0=\lambda V_0/2 + \lambda g M_s $ for the AFM1 state and $\varepsilon_{2,\lambda}^0=\lambda V_0/2 - \lambda g M_s$ for the AFM2 state under positive magnetic field, where $\lambda=+$ ($\lambda=-$) corresponds to the zLL on the top (bottom) surface. 
For Dirac surface states, besides the normal LLs, there are additional zeroth LLs (zLLs), depicted in Fig. \ref{fig:fig1}(c). All the higher LLs are equivalent for AFM1 and AFM2, 
%even in the presence of both electric and magnetic fields, 
and energy difference solely comes from zLLs with the eigen-energies $\varepsilon_{1,\lambda}^0=\lambda g M_s+ \lambda V_0/2$ for AFM1 and $\varepsilon_{2,\lambda}^0=\lambda g M_s- \lambda V_0/2 $ for AFM2 under positive $B$, where $\lambda=+$ ($\lambda=-$) corresponds to zLL on the conduction band bottom (valence band top), $V_0$ is asymmetric potential between two surfaces induced by electric fields, $g$ is the exchange coupling coefficient 
%between electrons and magnetic moments 
and $M_s$ is the saturation magnetization. 
%The energies of the zLLs depend on the signs of the magnetic gaps (the sign of $g$ and $m_{si,z}$ in $H_{e-M}$). 
When $g$ is positive, the zLL for the top surface state corresponds to the energy at the conduction band bottom (valence band top) for AFM1 (AFM2) while that for the bottom surface state is at the valence band top (conduction band bottom), in Fig. \ref{fig:fig1}(c). At $V_0=0$, the occupied zLL has the same energy for two AFM configurations $\varepsilon_{1,-}^0= \varepsilon_{2,-}^0=-g M_s$. 
%, where $g$ is the exchange coupling coefficient and $M_s$ is the saturation magnetization. %$\varepsilon_{1,-}^0=\varepsilon_{2,+}^0=-g M_s$,
However, this degeneracy will be broken by an electric field, which shifts the zLL energies $\varepsilon_{1,-}^0$ and $\varepsilon_{2,-}^0$ oppositely. For $V_0>0$, the zLL energy of AFM1 decreases ($\varepsilon_{1,-}^0=-g M_s-V_0/2$) while that of AFM2 increases ($\varepsilon_{2,-}^0=-g M_s+V_0/2$) (Fig. \ref{fig:fig1}(d)), and the energy difference is $\Delta\varepsilon =\varepsilon_{1,-}^0 -\varepsilon_{2,-}^0 =-V_0<0$. Therefore, AFM1 is energetically favored for $V_0>0$, corresponding to parallel alignment of electric and magnetic fields $\bm{B\cdot E}>0$ (we choose $V_0=eEL$ with $E$ representing electric field and $L$ the film thickness). For $V_0<0$ ($\bm{B\cdot E}<0$), AFM2 has a lower ground state energy ($\Delta\varepsilon=-V_0>0$). 
Therefore, the energy difference between two AFM states microscopically arises from the energy shift of zLLs under electric fields.

After identifying the lower-energy AFM configurations under magnetic and electric fields, we next study the sign of AH conductance. 
%in the two-surface-state model. 
%According to Fig. 3 in \emph{Zhao et al.} \cite{Zhao2021}, the MBT samples are heavily electron-doped so that the ordinary Hall resistance shows a negative slope. %For odd SL MBT films, when reducing magnetic fields from a large value to zero, the zero field AH resistance shares the same sign as the ordinary Hall resistance without any sign change. We find that only the Landau level spectrum for $g>0$ is consistent with the experimental observations within our current definition of the parameters, and thus we can fix the exchange coupling sign as positive (see Appendix Sec.I.B for more details). In contrast, the signs between zero field AH resistance and ordinary Hall resistance for even SL MBT films are opposite. 
%The sign of AH for each band is fixed by the magnetization vectors $m_{s1,z}$ and $m_{s2,z}$. We first consider the odd SL with $m_{s1,z}=m_{s2,z}=1$, which are in alignment with a positive magnetic field. 
To be consistent with negative AH sign for odd SLs in %\emph{Zhao et al.} 
Ref. \cite{zhao2021even}, exchange coupling $g$ should be positive, so that the valence bands of both surface states in odd SLs contribute negative AH sign (Appendix Sec.III \cite{sm2023}). For even SLs, the AH signs are reversed for the surface whose magnetization is flipped compared to that of odd SLs %Therefore, the valence band of the top (bottom) surface state should contribute a positive (negative) sign to the AH conductance for AFM1, and vice versa for AFM2, 
For Fermi energy inside magnetic gap, the valence bands of top and bottom surfaces give exact opposite contributions, leading to zero overall AH conductance, as shown in Fig. \ref{fig:fig1}(c) where blue and red colors stand for positive and negative AH signs, respectively. 
For the system with $B>0$ and $E>0$ ($V_0>0$) at electron doping, the favored AFM configuration is AFM1 with a positive AH conductance at electron doping (Fig. \ref{fig:fig1}(d)). When $E<0$ ($V_0<0$), the favored AFM configuration is AFM2, also exhibiting positive AH conductance at electron doping (Fig. \ref{fig:fig1}(e)). %Therefore, at electron doping, the energetically favored AFM state of even-SLs MBT films always possesses a positive sign of the AH conductance under a positive magnetic field, regardless of the direction of the electric field. 
%Under negative external magnetic field, similar analysis suggests that the odd SL MBT film displays positive AH, while the even SL shows negative AH. 
With similar analysis for $B<0$, we conclude that the odd and even SL films will always have opposite AH signs at electron doping, independent of the alignment between $E$ and $B$. These results explain the observations of the even-odd AH effect in the samples with electron doping.

%\addCXL{We may move this paragraph to appendix if it is still too long. }
%The situation is dramatically changed for hole doping.
%For $B>0$ and $E>0$, the Fermi level now crosses the valence band of the bottom surface in AFM2, so that all the occupied states contribute a negative AH conductance, which is the same sign of the AH conductance in odd SL. Similar situation occurs for $B>0$ and $E<0$. 
%The odd SL and even SL both exhibit negative (positive) AH sign under positive (negative) magnetic field, again regardless of the electric field direction. This is consistent with the experiment in \emph{Gao et al.} \cite{Gao2021}, where the 5-SL MBT exhibits negative AH at both electron and hole doping, while the 6-SL MBT displays negative and positive AH resistance at hole and electron doping, respectively.

\begin{figure}
\includegraphics[width=\columnwidth]{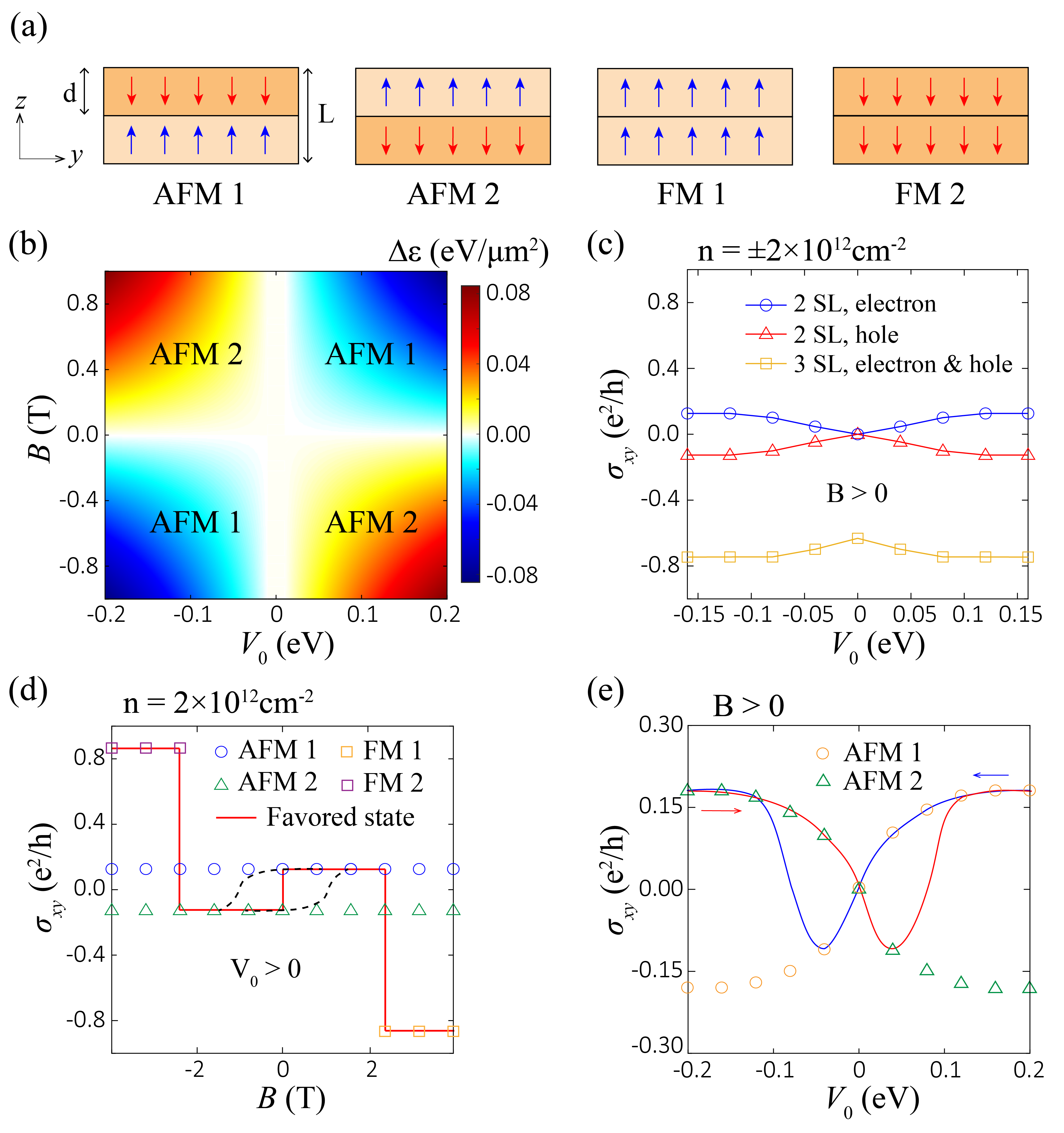}
\caption{(a) Illustration of thin film model for 2 SL MBT films. Each SL has a thickness of $d$. %For AFM states, the adjacent layers have anti-parallel magnetization and for FM states the two layers have parallel magnetization. 
(b) The energy density difference between AFM1 and AFM2 $\Delta \varepsilon =\varepsilon_1 -\varepsilon_2$ as a function of asymmetric potential $V_0$ and magnetic field $B$ at electron density $n=2\times 10^{12} cm^{-2}$. (c) Numerically calculated Hall conductance $\sigma_{xy}$ as a function of $V_0$ for favored 2 SL and 3 SL samples at electron and hole doping under positive $B$ with carrier density $n=\pm 2\times 10^{12} cm^{-2}$. (d) Hall conductance $\sigma_{xy}$ as a function of $B$ at electron density $n=2\times 10^{-12} cm^{-2}$ for positive $V_0$. The red line is the expected favored state at different $B$. %The spin-flop transition field is around 2.3 T for even MBT films. 
The dashed black lines illustrate the hysteresis loop. (e) Electric control of Hall conductivity for 2 SL. The yellow circles and green triangles stand for the Hall conductivity for AFM1 and AFM2 respectively.  The solid lines of sweeping $V_0$ are sketched by hand only for illustration.}
\label{fig:fig2}
\end{figure}

To buttress our arguments, we also investigate a thin film model which includes both surface and bulk states, and perform numerical calculations for the zLL energies and the AH conductivity for 2-SL MBT in Fig. \ref{fig:fig2} (Appendix Sec.II.B \cite{sm2023}). %Although we only show the results for 2 SL MBT film for simplicity, the conclusion can be applied to thicker even SL films. For 2 SL MBT, there are four possible AFM and FM configurations as shown in Fig. \ref{fig:fig2}(a). We denote the ground state energy density of the AFM1 and AFM2 configurations as $\varepsilon_1$ and $\varepsilon_2$, respectively, and calculate the energy density difference $\Delta \varepsilon=\varepsilon_1-\varepsilon_2$ as a function of the magnetic field $B$ and the asymmetric potential $V_0=-eEL$ at electron density $n= 2\times 10^{-12} cm^{-2}$ induced by the out-of-plane electric fields $E$ with $L$ being the film thickness (Fig. \ref{fig:fig2}(b)). 
For $\bm{B\cdot E}>0$, $\Delta \varepsilon=\varepsilon_{AFM1}-\varepsilon_{AFM2}<0$ so that AFM1 is energetically favored, while for $\bm{B\cdot E}<0$, AFM2 is preferred (Fig. \ref{fig:fig2}(b)). 
We then compare the Hall conductivity in 2-SL and 3-SL MBT at both electron and hole doping with carrier density $n=\pm 2\times 10^{-12} cm^{-2}$ for positive $B$ in Fig. \ref{fig:fig2}(c).
%For 3-SL MBT films, the top and bottom SLs have parallel magnetization, and the middle SL has opposite magnetization due to AFM (see Appendix Fig. S2(a)). At the asymmetric potential $V_0=0$, the 2 SL MBT presents zero Hall conductance as the top and bottom surface states cancel out their contributions, while the 3 SL MBT has a negative AH conductance for both electron and hole-doping. 
A non-zero $V_0$ can induce a positive (negative) AH response in 2-SL MBT at electron doping (hole doping). For 3-SL MBT, the AH sign does not change with doping or electric fields. 
%On the other hand, the doping does not change the sign of AH conductance in 3-SL MBT, which only weakly varies with electric fields. 
In summary, the 2-SL and 3-SL MBT show opposite AH signs at electron doping, while they have the same AH sign at hole doping, consistent with the analysis of the two-surface-state model.  This prediction is in agreement with the experiment in Ref. \cite{gao2021layer},  
%\addCXL{Move the following to Appendix. }, where it was observed that 5-SL and 6-SL MBT films show opposite AH signs at electron doping and the same AH sign at hole doping. We also estimate the energy difference $\Delta \varepsilon/n$ to be around $0.1meV$ for an magnetic field of $1T$ and electric field of $0.2V/nm$ ($\Delta\varepsilon=1eV/\mu m^2$ and $n=1\times 10^{12}cm^{-2}$), which is at the same order as the temperature in experiments (around 2K). %Add the energy estimate and say it is consistent with Gao's experiment. 
and the resulting AH sign is also consistent with previously proposed even-odd effect in the insulating regime \cite{otrokov2019unique,li2019intrisic,zhang2019topological} (Appendix Sec.III \cite{sm2023}).

We further investigate the hysteresis loop for even SL MBT. Fig. \ref{fig:fig2}(d) shows the Hall conductivity $\sigma_{xy}$ for both the AFM and FM states at electron density $n=2\times 10^{-12} cm^{-2}$ for $V_0>0$ and sketch the expected favored states at different $B$ by red line. 
%To determine the favored state between FM and AFM states, we refer to the magnetization energy $E_M$. 
%We estimate the spin-flop transition occurring around the field $B_c^{\pm}=\pm B_c\approx \pm J/M_s$, (as the in-plane magnetization is not included here, we did not study the canted AFM phase during the spin-flop transition). 
The spin-flop transition field $B_c$ for 2-MBT film is around $2.3$ T in experiment \cite{zhao2021even}. %\addCXL{it seems we have not defined what is FM1 and FM2, maybe we should label it in Fig. 1.  } 
For $|B|>B_c$, the FM states have the lower magnetization energy and thus energetically favored. 
%the magnetization of both SLs tends to align with the magnetic field. 
%, which means that FM1 is favored at $B>B_c^+$ and FM2 is favored at $B<B_c^-$. 
%\addCXL{ $B_c^{\pm}$ has not been defined. }
For $|B|<B_c$, the AFM states have lower energy, and AFM1 is favored at $0<B<B_c$, while AFM2 is preferred at $-B_c<B<0$ for $V_0>0$. 
%As shown in Fig.  \ref{fig:fig1}(a), the typical value of the spin-flop field $B_c$ for 2 SL MBT film in the experiment is around $2.3$ T \cite{Zhao2021}. 
When the magnetic field is swept from positive to negative, the favored state for 2-SL films goes through FM1 $\rightarrow$ AFM1 $\rightarrow$ AFM2 $\rightarrow$ FM2 and correspondingly the sign of Hall conductivity $\sigma_{xy}$ varies as $-\rightarrow +\rightarrow - \rightarrow +$ (the red lines in Fig. \ref{fig:fig2}(d)). Since the AFM1-AFM2 phase transition is of first order, a hysteresis loop can form at small $B$ before spin-flop transition (the black dashed lines in Fig. \ref{fig:fig2}(d)), corresponding to the observed AH hysteresis loop in Fig. \ref{fig:fig1}(a). Thus, the double sign changes of the hysteresis loop can be naturally understood as a two-step phase transition: the first-step transition between two AFM states followed by the second-step transition between the AFM and FM states. For even SL MBT films thicker than 2 SL, multi-step spin-flop phase transitions might occur due to multiple FM configurations \cite{yang2021odd} (Appendix Sec.IV \cite{sm2023}).

The dependence of AFM ground state energy on electric fields implies the possibility of electrical control of AH conductance near AFM transition. %For a positive magnetic field, the Hall conductance of the AFM1 (AFM2) configuration is positive (negative) for $V_0>0$, negative (positive) for $V_0<0$ and zero at $V_0=0$. 
If we sweep electric fields from positive to negative (the blue curve in Fig. \ref{fig:fig2}(e)), the favored configuration changes from AFM1 to AFM2 according to Fig. \ref{fig:fig2}(b), and correspondingly, Hall conductivity first changes from positive to negative momentarily then back to positive due to hysteresis.
% so the Hall conductance of the favored state should always be positive. However, due to hysteresis, the system stays in AFM 1 for a short moment after the electric field flips before transiting to AFM 2. As a result, the Hall conductivity first changes from positive to negative momentarily then back to positive, so the sign change in AH due to the hysteresis loop is an indicator of the transition between two AFM states. 
This electric control of AH conductance potentially provides a microscopic picture to understand recent experiments
%on the AH sign by scanning the electric field back and forth 
\cite{gao2021layer}. 
%This observation was interpreted as the consequence of the phenomenological axion electrodynamics while our understanding based on the zLLs is equivalent to that but provides a microscopic physical picture. 

%\section*{Orbital magnetization and magnetoelectric effect in MBT films}

{\it Orbital magnetization and magnetoelectric effect - }
Next we will discuss an alternative view point of AFM transition based on orbital magnetization created by magnetoelectric effect in MBT films. 
%Magnetic moments have two origins: spin and orbital moment. Normally,
In magnetic materials, spin moment is usually much larger than orbital moment. The odd-SL MBT has an uncompensated net spin magnetization, and thus orbital magnetization is negligible. For even SLs, however, spin magnetization cancels out in AFM configurations, and orbital magnetization can play a role. The orbital magnetization in even SL MBT can lead to magnetoelectric effect, e.g. an electric field can create a magnetization, given by ${\bf M}=\alpha {\bf E}$, with magnetoelectric coefficient $\alpha$. The magnetoelectric effect has been previously studied in 2D magnetic materials \cite{lu2020artificial,shang2021robust}. 
%As studied below, %in the MBT films, this electric field induced orbital magnetization depends on the AFM configurations and has opposite sign between AFM1 and AFM2, so that it 
%the orbital magnetization can provide an alternating understanding on different energies between two AFM configurations. 

% The orbital magnetization usually contains two parts, a trivial part and a topological part \cite{Xiao2005,Souza2008,Xiao2010,Thonhauser2011},
% \begin{equation}
%     \begin{gathered}
%     m_{total}=m_{trivial}+m_{topological}, \\
%     m_{trivial}=-\frac{ie}{2\hbar} \langle \nabla_k u| \times [H(k)-\epsilon(k)]| \nabla_k u\rangle, \\
%     m_{topological}=-\frac{ie}{2\hbar} \langle \nabla_k u| \times 2 [\epsilon(k)-\mu]|\nabla_k u \rangle,
%     \end{gathered}
% \end{equation} 
% where $\epsilon(k)$ and $|u\rangle$ are the eigenvalues and eigenstates of the Hamiltonian $H(k)=H_e+H_{e-M}$ of the system, and $\mu$ is the Fermi energy. Intuitively, the trivial part comes from the self rotation of a wave packet around its center of mass, and the topological part originates from the center of mass motion under the presence of boundary potentials, which relates to the Berry curvature.

\begin{figure}
\includegraphics[width=\columnwidth]{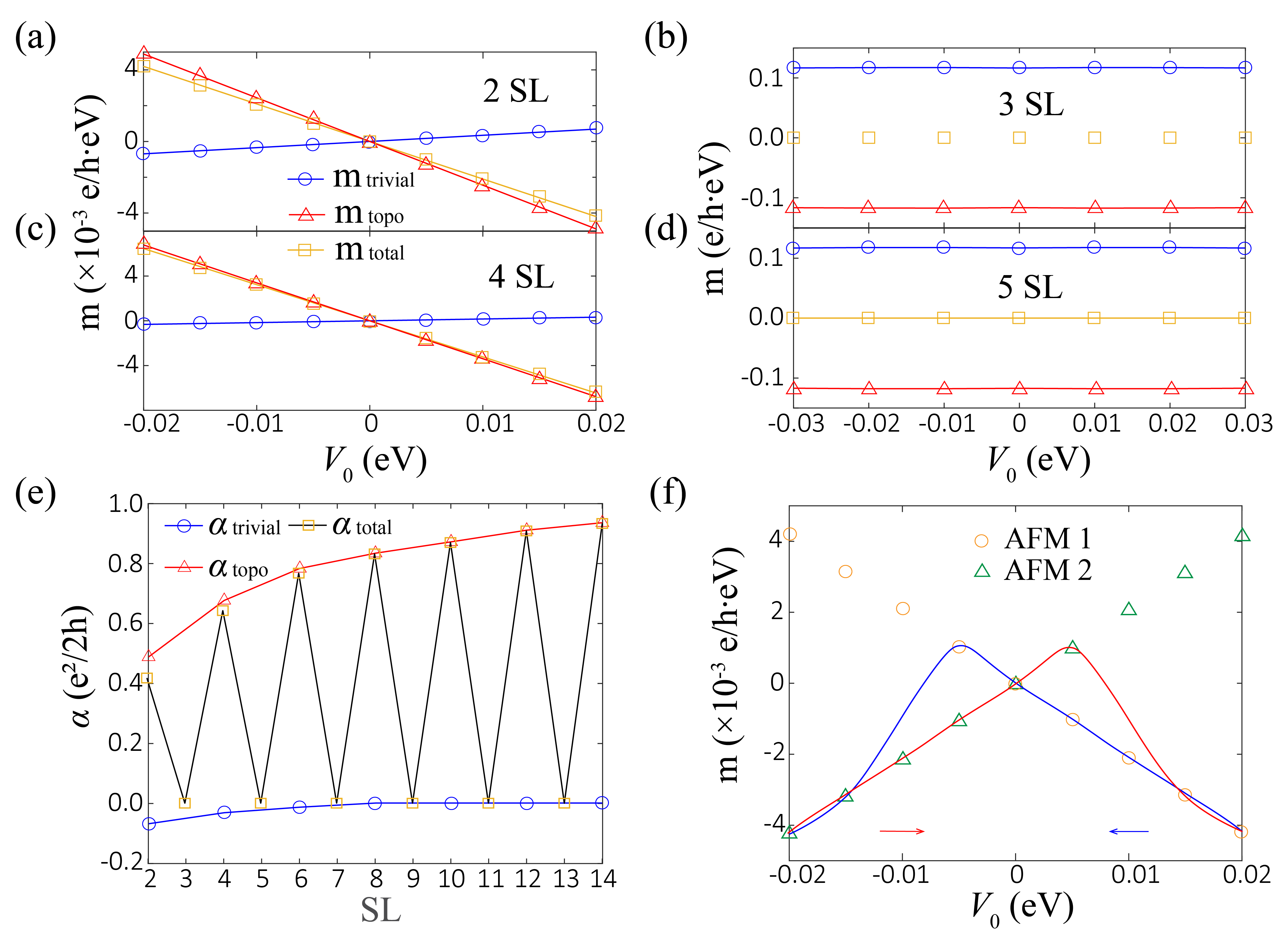}
\caption{Calculated orbital magnetic moment $m$ as a function of $V_0$ for (a) 2SL, (b) 3SL, (c) 4SL and (d) 5SL in the thin film model for chemical potential $\mu=0$. The blue, red and yellow lines are for trivial, topological and total magnetic moment, respectively. %In even SLs, $m$ displays non-zero slope versus $V_0$, while in odd SLs, $m$ is a constant under varying $V_0$. 
(e) The trivial and topological part of $\alpha$ and total $\alpha$ as a function of SL number. (f) Illustration of the electric control of orbital magnetic moment in even SL MBT. }
\label{fig:fig3}
\end{figure}

Orbital magnetization usually contains two parts, a trivial and a topological part, $\bm{m}_{total}=\bm{m}_{trivial}+\bm{m}_{topo}$ %, with different physical origins
\cite{xiao2005berry,souza2008dichroic,xiao2010berry,thonhauser2011theory}. Fig. \ref{fig:fig3}(a)-(d) show orbital magnetic moments as a function of $V_0$ for 2-5 SLs in thin film model at $\mu=0$ (Appendix Sec.VII \cite{sm2023}). For odd SLs (Fig. \ref{fig:fig3}(b), (d)), orbital moment remain constants, while for even SLs (Fig. \ref{fig:fig3}(a), (c)), orbital moment is zero at $V_0=0$ and linearly increases with $V_0$ for AFM1. 
%Fig. \ref{fig:fig3}(a), (c) show the orbital magnetic moment for the 2 and 4 SLs in AFM1 configuration, which is zero for $V_0=0$ and linearly increases with $V_0$. 
This linear behavior implies the magnetoelectric effect in even SL MBT. The signs of magnetic moments reverse for AFM2. With magnetic fields, the electric-field-induced orbital magnetization can lead to the energy difference $\Delta \varepsilon=\varepsilon_{AFM1}-\varepsilon_{AFM2}=\bm{M}_{orb,1}\cdot \bm{B}-\bm{M}_{orb,2}\cdot \bm{B}$ between two AFM states, where $\bm{M}_{orb,1}$ and $\bm{M}_{orb,2}$ are orbital magnetization of AFM1 and AFM2, respectively. A positive electric field, namely $V_0>0$, can induce a negative (positive) total orbital magnetization $\bm{M}_{orb,1}=-M_0\hat{z}$ ($\bm{M}_{orb,2}=M_0\hat{z}$) for the AFM1 (AFM2) configuration, where $M_0$ is a positive number and $\hat{z}$ is the unit vector along z axis, so $\Delta \varepsilon=-2M_0B<0$ at positive $B$, which means AFM1 is favored, and $\Delta \varepsilon=2M_0B>0$ at $-B$, for which AFM2 is favored. The above analysis is quantitatively consistent with the perspective of zLLs (Appendix Sec.VII \cite{sm2023}). 
%namely AFM1 is the favored state for $\bm{B\cdot E}<0$ and vise versa. 

%In even SLs, both trivial and topological parts have non-zero slopes while they remain constants in odd SLs, consistent with the results for two-surface state model (see Appendix Sec.III). 
%The magnitude of the calculated orbital magnetization is estimated and compared to the Bohr magneton. 
For an electric field strength $E\approx 0.1V/nm$, orbital moments in even SL MBT is estimated as $10^{-1} \frac{e}{h}\cdot eV \sim 0.4 \mu_B/nm^2$ with $ \frac{e}{h}\simeq 4.18 \mu_B/(nm^2\cdot eV)$ and Bohr magneton $\mu_B=\frac{e\hbar}{2m_e}$. 
%=\frac{2\pi e}{h}R_y a_0^2$, Rydberg constant $R_y \simeq 13.6eV$ and Bohr radius $a_0 \simeq 5.29\times 10^{-11}m$. 
With the magnetic moment $\sim 5 \mu_B$ of Mn ions and the in-plane lattice constant $a\simeq 0.43 nm$ of MBT \cite{yan2019crystal}, spin magnetization is around $27 \mu_B/nm^2$, and hence orbital magnetization is approximately two orders smaller than spin magnetization, which thus can only play a role in compensated AFM configuration.

We can further extract the magnetoelectric coefficient $\alpha$ from orbital magnetization \cite{essin2009magnetic,essin2010orbital,wang2015quantized}. %defined by $\bf{M}=\alpha \bf{E}$.
In Fig. \ref{fig:fig3}(e), the trivial part of $\alpha$ goes to zero and the topological part approaches quantized value $e^2/2h$ as the layer number increases for even SLs at $\mu=0$ \cite{li2022identifying,chen2023side}.
The odd SLs always exhibit zero $\alpha$ with a constant orbital magnetization. Thus, $\alpha$ oscillates between zero and nonzero for odd and even SLs of MBT films (Fig. \ref{fig:fig3}(e)). The behaviors of $\alpha$ for a nonzero chemical potential $\mu$ are discussed in Appendix Sec.VII \cite{sm2023}.

We should distinguish magnetoelectric coefficient $\alpha$ from the axion parameter $\theta$, a three-dimensional (3D) bulk quantity that characterizes the axion term $\theta e^2 {\bf E\cdot B}/2\pi h$ in electromagnetic response of topological insulators \cite{wilczek1987two, wang2015quantized, essin2009magnetic, essin2010orbital, zhang2019topological, qi2008topological, sekine2021axion}. 
%In the field theory description of 3D TIs, a topological $\theta$ term $\theta e^2 {\bf E\cdot B}/2\pi h$ is added into the ordinary Maxwell electromagnetic Lagrangian, in which %${\bf E}$ and ${\bf B}$ are the conventional electric and magnetic fields inside an insulator, and $\theta$ is a pseudo-scalar and called the axion parameter. 
%When time reversal symmetry is present, $\theta$ can only takes two quantized values 0 and $\pi$, corresponding to trivial insulators and TIs, respectively \cite{qi2008topological}. The axion parameter 
$\theta$ can be directly connected to magnetoelectric coefficient $\alpha$ (an experimental observable) as $\alpha = \theta e^2/2\pi h $ when all the surface states of 3D TIs are gapped. %This can only be achieved in magnetic TIs with the AFM alignment of magnetization between top and bottoms surfaces, so that all surface states are gapped out. 
For the MBT films, this corresponds to even SLs in the large thickness limit, and the magnetoelectric coefficient $\alpha$ value approaches $e^2/2 h$ (Fig. \ref{fig:fig3}(e)) as $\theta = \pi$ in bulk MBT. %, namely the topological magnetoelectric effect for the axion insulator phase. 
For the thick odd-SL MBT, magnetoelectric coefficient $\alpha$ is zero, different from the bulk axion parameter $\theta=\pi$. Due to $\theta=\pi$, such phase was previously referred as AIs with higher-order topology 
\cite{xu2019higher,zhang2020mobius,wieder2018axion,yue2019symmetry,varnava2018surfaces,chen2021using,varnava2021controllable}. 
%, in which chiral modes propagate along the hinges. 
 In Appendix Sec.VII \cite{sm2023}, we show while the total orbital magnetization depends on bulk magnetic configurations, the magnetoelectric coefficient $\alpha$ only depends on the surface magnetization, insensitive to bulk magnetization. 

%({\bf \color{red} RM: further simplify this paragraph?})

%Even though the spin magnetization is zero in even SLs, the 
Orbital magnetization in even SL MBT can have an impact in magnetic circular dichroism (MCD) \cite{qiu2023axion}. Early RMCD experiments in even SLs show a non-zero hysteresis loop around small magnetic fields
%, presumably coming from magnetic domains/disorders 
\cite{ovchinnikov2021intertwined,yang2021odd}. Our studies of orbital magnetization provides an intrinsic mechanism for these observations. 
%A decent RMCD signal may also come from the p-d transition of magnetic ions while the orbital magnetization discussed here may be more sensitive to a small photon energy that matches the topological gap. Thus, examining the frequency dependence of RMCD may provide information of the origins of RMCD signals. We also notice that a recent work suggests the RMCD signals for even-SL MBT can exist as the reflections from the top and bottom surface states are not identical, leading to a finite Kerr rotation at a certain photon wavelength in the Kerr experiment configuration, while the transmission part of the MCD (Faraday rotation) remains zero \cite{ahn2022theory}. Therefore, while the RMCD signals may have the contributions from both mechanisms, the transmission MCD signals will be dominated by orbital magnetization discussed here. 
A decent RMCD signal may also come from the p-d transition of magnetic ions, or the difference in the reflections between two surface states \cite{ahn2022theory}. MCD signals of orbital magnetization in even SLs can also be controlled by sweeping electric fields. Following the blue curve in Fig. \ref{fig:fig3}(f), AFM1 is the favored configuration with negative orbital magnetization at positive $V_0$, and magnetization vanishes at $V_0=0$. As $V_0$ turns to negative, the system remains in the AFM1 with positive orbital magnetization as the AFM1-AFM2 transition is of first order, giving rise to hysteresis loop, similar to AH hysteresis loop discussed in Fig. \ref{fig:fig2}(e). %The transition from AFM1 to AFM2 is expected to occur at a certain negative value of $V_0$, and the orbital magnetization changes to negative after this value of $V_0$. 
Therefore, orbital magnetization is expected to vary from negative to positive then back to negative as the electric potential $V_0$ sweeps from positive to negative.

%\section*{Conclusion and discussion}
{\it Conclusion - }
In summary, we apply a two-surface-state model and a thin-film model to MBT films, and demonstrate that the presence of electric and magnetic fields can select a favored AFM configuration in even SLs through the effect of zLLs, leading to a nonzero AH response and orbital magnetization. %Our results provide a possible explanation of hysteresis loop forms for even and odd SL MBT. 
For real samples, disorders and magnetic domains are inevitable (Appendix Sec.VI \cite{sm2023}). For example, antiferromagnetic domain walls have been imaged in MBT via cryogenic magnetic force microscopy \cite{sass2020magnetic}. Thus, the AFM transition should correspond to enlargement and shrinkage of two opposite AFM domains. Furthermore, bulk states, in addition to surface states, may also play a role due to chemical potential inhomogeneity, potentially leading to more complicated behaviors \cite{zhang2020experimental}. 
%, and our prediction here is more applicable to high-quality samples with low carrier concentrations. 
%We further propose that orbital magnetization induced by electric field in even SL can result in nonzero MCD signals in the AFM regime. 
More experimental studies are necessary to validate our prediction of AH effect in even and odd SL MBT at electron and hole dopings, as well as the possible electric control of orbital magnetization in even SL MBT.

%\section*{Acknowledgement}
{\it Acknowledgement - }
We would like to acknowledge Binghai Yan for the helpful discussion. R.B. M. and C.-X. L. acknowledges the support through the Penn State MRSEC–Center for Nanoscale Science via NSF award DMR-2011839.  C.-X. L. also acknowledges the support for the NSF grant award (DMR-2241327). Y.-F. Z. and C.-Z. C. acknowledge the support from the ARO Award (W911NF2210159) and the Gordon and Betty Moore Foundation’s EPiQS Initiative (Grant GBMF9063 to C. -Z. C.). D.X. is supported by AFOSR MURI 2D MAGIC (FA9550-19-1-0390).

\bibliography{ref.bib}

\clearpage

% Supplemental materials

\pagebreak
\widetext
\begin{center}
	{\huge{\bf Supplemental Materials}}
\end{center}

\setcounter{equation}{0}
\setcounter{figure}{0}
\setcounter{secnumdepth}{2}
\renewcommand{\theequation}{S\arabic{equation}}
\renewcommand{\thefigure}{S\arabic{figure}}
\renewcommand{\thetable}{S\arabic{table}}
\renewcommand{\appendixname}{Supplement}

\section{A summary of experimental observations in MBT films}
In this section, we summarize our experimental observations of the anomalous Hall transport in MnBi$_2$Te$_4$ films, which was published in Ref. \cite{zhao2021even}, make the manuscript self-contained. We used molecular beam epitaxy (MBE) to fabricate the MnBi$_2$Te$_4$ films down to 1SL and systematically studied their thickness-dependent transport properties. In MnBi$_2$Te$_4$ films thicker than 2 SL, we identified a non-square hysteresis loop in the AFM regime. Furthermore, the non-square AH hysteresis loop observed in the AFM state shows an even-odd layer number dependent behavior. In even SL samples, hysteresis loops present a hump feature, while in odd SL samples, it shows a two-step magnetic transition feature. We speculated that this feature may be induced by the superposition of two AH effects with opposite and same signs for the even and odd layer numbers, respectively. To extract these two AH components, we expressed the $\rho_{yx}$ of the MnBi$_2$Te$_4$ films as: $\rho_{yx} = \rho_{OH} + \rho_{AH1} + \rho_{AH2}$ and fitted the $\rho_{yx}'=\rho_{yx}-\rho_{OH}$ with the following equation:
\beq
    \rho_{yx}'=\rho_1 tanh(w_1(\mu_0H-\mu_0 H_{C1})) +\rho_2 tanh(w_2(\mu_0H-\mu_0 H_{C2})) 
\eneq
Here $\rho_{OH}$ represents the ordinary Hall effect, $\rho_1$($\rho_2$), $\mu_0 H_{C1}$($\mu_0 H_{C2}$) are the amplitude and the coercive field of the first (second) AH component $\rho_{AH1}$ ($\rho_{AH2}$), respectively; $w_1$ and $w_2$ are two constants. The extracted AH1 possesses a more pronounced coercive field and displays a distinct even-odd layer-dependent behavior, which is the central theme of this paper. AH2 has a smaller coercive field and always shows a negative sign in all the samples. Through a careful analysis, we found that the two AH components originate from two coexisting phases in the MBE-grown MnBi$_2$Te$_4$ films: AH1 is from the dominant MnBi$_2$Te$_4$ phase while AH2 is from the minor Mn-doped Bi$_2$Te$_3$ phase. 

\begin{figure}
\includegraphics[width=\textwidth]{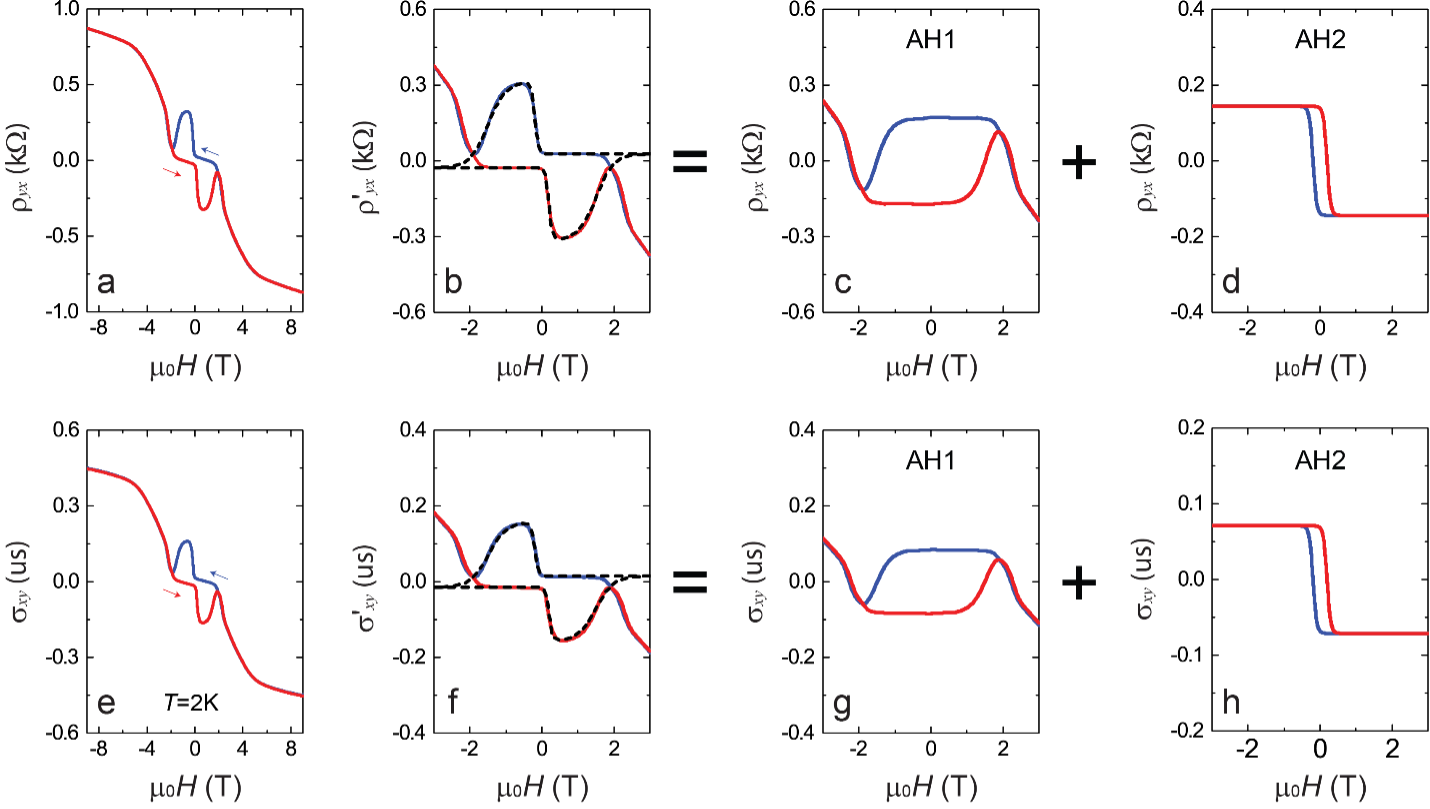}
\caption{Hall trace of MBE-grown 2SL MnBi$_2$Te$_4$ films measured at T =2 K. (a) 
 $\mu_0 H$ dependent $\rho_{yx}$ (a) of 2SL MnBi$_2$Te$_4$ films. (b) The corresponding AH resistance $\rho_{yx}'$ after subtracting the ordinary Hall effect in the AFM regime. The dashed lines are curves fitted by the two AH component model in Ref. \cite{zhao2021even}. (c-d) The extracted first (c) and second (d) AH component $\rho_{AH1}$ and $\rho_{AH2}$. The blue (red) curve represents the process for decreasing (increasing) $\mu_0 H$. (e-h) The corresponding AH conductance of (a-d). }
\label{fig:Exp}
\end{figure}

In Fig. \ref{fig:Exp}, we show both the AH resistivity and conductivity fitting results of the 2SL MnBi$_2$Te$_4$ film. To derive the AH conductivity, we first converted the Hall resistivity $\rho_{yx}$ (Fig. \ref{fig:Exp}a) of the 2SL MnBi2Te4 film into conductivity $\sigma_{yx}$ (Fig. \ref{fig:Exp}e) using the formula $\sigma_{yx}=\rho_{yx}/(\rho_{yx}^2+\rho_{xx}^2)$. We then employed the same fitting method for AH conductivity as we did for AH resistivity. The extracted two AH conductivity components are shown in Fig. \ref{fig:Exp}g and h , mirroring the characteristics observed in the AH resistivity. 

%{\color{blue} Add the experiment description. Revise the following paragraphs. Add the Hall conductivity data and the corresponding figures. }

\section{Model Hamiltonian for MBT films} \label{sec:Hamiltonian_MBTfilms}
\begin{figure}
\includegraphics[width=\textwidth]{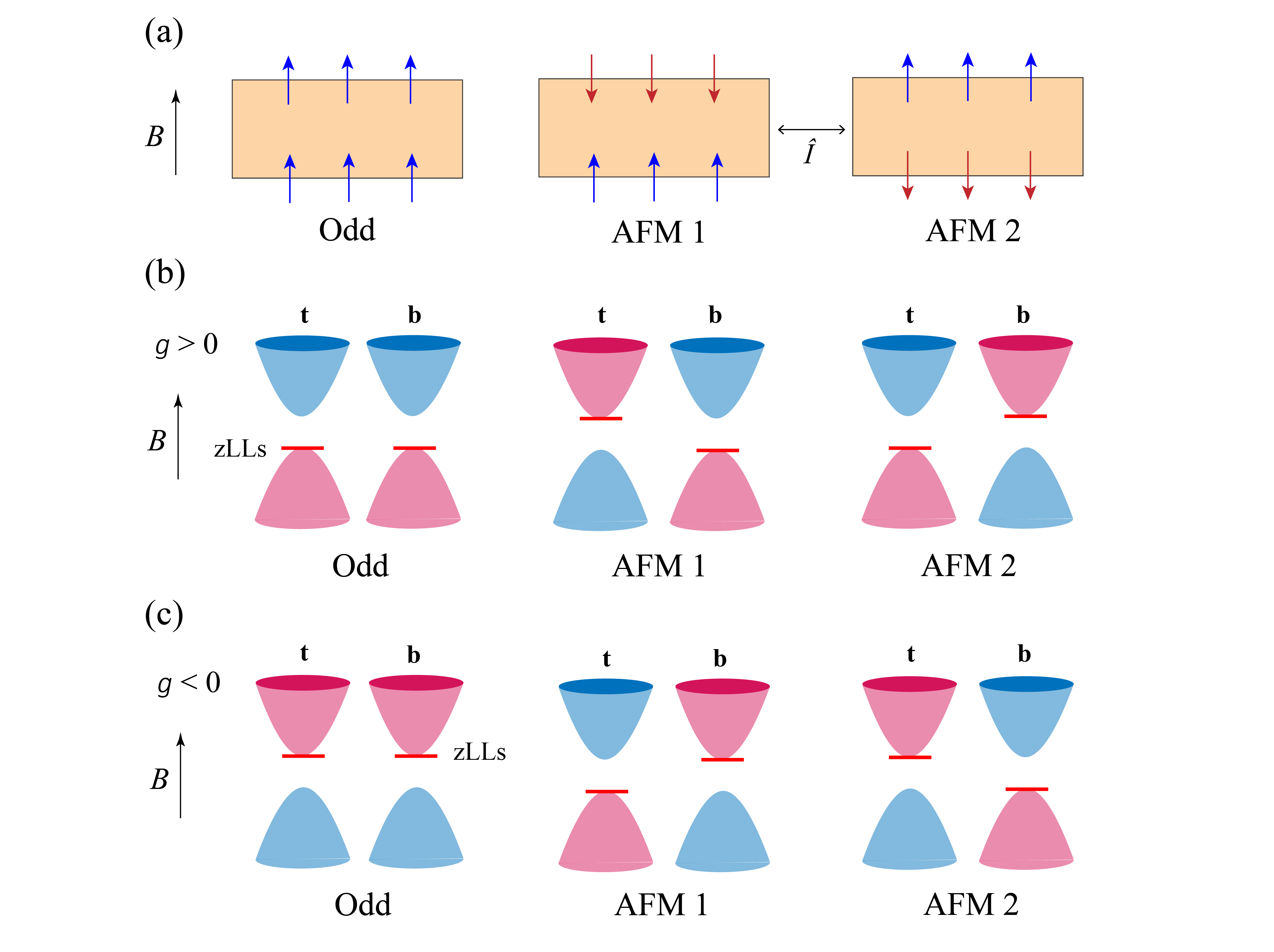}
\caption{(a) Illustration of the magnetization configurations for odd and even SL AFM MBT films. In odd SL, the magnetization of both surfaces aligns with the external magnetic field. In even SL, there are two possible AFM configurations with opposite magnetization on top and bottom surfaces. (b) The positions of zLLs in two-surface-state model for $g>0$ and $B>0$. The bands are labeled with red and blue colors which represent negative and positive AH signs, respectively. (c) The locations of zLLs for $g<0$ and $B<0$.}
\label{fig:figS1}
\end{figure}

MnBi$_2$Te$_4$ (MBT) is a layered topological insulator with anti-ferromagnetic (AFM) coupling between adjacent septuple layers (SLs). Here we use two models to simulate such MBT films. In a two-surface-state model, we only consider the top and bottom surfaces of the film described by two Dirac Hamiltonians. We further introduce exchange couplings, magnetization, gauge couplings and external magnetic fields, as well as an asymmetric potential induced by external electric fields (gate voltages). The AFM order implies that the magnetizations on the two surfaces are parallel for odd SL MBT films and anti-parallel for even SL films (Fig. \ref{fig:figS1}(a)). In a thin-film model, we employ an effective four-band model for a bulk magnetic topological insulator and construct a slab model for the thin film along $z$ direction with the open boundary condition for the four-band model Hamiltonian (Fig. \ref{fig:figS2}(a)). The exchange coupling is a $z$-dependent function that implements the AFM order between neighboring layers. We further investigate the Landau levels and anomalous Hall (AH) effect for both models in the following sections.

\subsection{Two surface states model} \label{sec:two_surface_model}
As described in the main text, the two-surface-state model writes as
% \begin{equation}\label{eq:two_surface}
%     H=v_f((k_x+\frac{e}{\hbar} A_x) \sigma_y-(k_y+\frac{e}{\hbar}  A_y) \sigma_x)\tau_z+V_0\tau_z/2+g M_s \begin{pmatrix}
% m_{1,z} \sigma_z & 0 \\
% 0 & m_{2,z}\sigma_z
% \end{pmatrix},
% \end{equation}
% where $\tau$ and $\sigma$ are Pauli matrices in the layer and spin sub-space, respectively, $m_{i,z}=\pm 1$ are the magnetization vectors of the top ($i=1$) and bottom ($i=2$) surfaces (in even SL, $m_{1,z}=-1$ corresponds to AFM1, $m_{1,z}=+1$ corresponds to AFM2 and $m_{2,z}=-m_{1,z}$), and $g$ is the exchange coupling coefficient. Below we will solve the Landau level spectrum of this model Hamiltonian. 
\beq 
H=H_M+H_e+H_{e-M}.
\eneq
Specifically, 
\beq H_M=J \bm{m}_{s1}\cdot \bm{m}_{s2}- M_s \bm{B} \cdot (\bm{m}_{s1}+\bm{m}_{s2}) - \frac{K}{2} (m_{s1,z}^2+m_{s2,z}^2),
\eneq 
where $\bm{m}_{si}=(sin\theta_i cos\phi_i, sin\theta_i sin\phi_i, cos\phi_i)$ is the magnetization vector with polar and azimuthal angles $\theta$ and $\phi$, the index $i=1,2$ labels the magnetization at the top and bottom surfaces, $M_s$ is the saturation magnetization, $J$ labels the effective exchange coupling between magnetizations at two surfaces ($J>0$ for AFM and $J<0$ for FM), $\bm{B}$ is the external magnetic field and $K$ is easy-axis anisotropy \cite{li2022identifying,mills1968surface,ovchinnikov2021intertwined}. The magnitude of magnetization has been absorbed into the definition of the parameters $J$, $K$ and $M_s$, so $\bm{m}_{si}$ is a unit vector ($|\bm{m}_{si}|=1$).  
$H_e$ is described by two-surface-state model in the basis $|t,\uparrow\rangle$, $|t,\downarrow\rangle$, $|b,\uparrow\rangle$ and $|b,\downarrow\rangle$,
\beq H_e=v_f((k_x+\frac{e}{\hbar}A_x) \sigma_y-(k_y+\frac{e}{\hbar}A_y) \sigma_x)\tau_z+ V_0\tau_z/2, \eneq
where $\sigma$ and $\tau$ are Pauli matrices in the spin and layer sub-space, $v_f$ is the Fermi velocity, $V_0$ is the asymmetric potential between two surfaces and the Landau gauge is chosen as $\bm{A}=(0,A_y,0)=(0,Bx,0)$ for the out-of-plane magnetic field $\bm{B}=(0,0,B)$. The electron-magnetization coupling Hamiltonian depends on magnetic configurations, and takes the form
\beq H_{e-M}=g M_s \begin{pmatrix}
m_{s1,z} \sigma_z & 0 \\
0 & m_{s2,z}\sigma_z
\end{pmatrix}
\eneq
where $g$ is the exchange coupling coefficient between electrons and magnetic moments and the two blocks of the above Hamiltonian are for two surfaces. Here the exchange coupling of in-plane magnetization is dropped because it only shifts the locations of surface Dirac point and can be generally absorbed into the gauge potential ${\bf A}$. The directly Zeeman coupling between electron spin and external magnetic field is dropped as it is much smaller than the exchange coupling at the low magnetic field limit. The ground state of $H_M$ is given by the out-of-plane AFM configurations with $\bm{m}_{s1}=(0,0,\pm1)$ and $\bm{m}_{s2}=-\bm{m}_{s1}$ 
at zero or low external magnetic fields $B$ \cite{zhang2019topological,ovchinnikov2021intertwined} ($m_{s1,z}=-1$ corresponds to AFM1, $m_{s1,z}=+1$ corresponds to AFM2). The magnetization energy for these two AFM states is $E_M=-J-K$, and the energy of AFM states is independent of magnetic field $B$ due to the zero total magnetization $\bm{m}_{s1}+\bm{m}_{s2}=0$. On the other hand, the FM state has the magnetization energy $E_M=J-K\pm 2M_sB$, where $\pm$ selects the FM state with magnetization vectors aligned with $B$ as the favored configuration. Thus, for $J>0$, the AFM states have lower energy while the FM state can be energetically favored at a large $B$. 

We next will solve the Landau level problem for the above two-surface-state model. Under the presence of an out-of-plane magnetic field $\mathbf{B}=(0,0,B)$, Landau levels (LLs) emerge from the bands \cite{novik2005band}. We choose the Landau gauge as $\mathbf{A}=(0,Bx,0)$ with $B>0$ and substitute $k_x$ and $k_y$ with the annihilation and creation operators 
\begin{equation}
    a=\frac{l_c}{\sqrt{2}}k_-, a^{\dagger}=\frac{l_c}{\sqrt{2}}k_+,
\end{equation}
where $k_{\pm}=k_x \pm ik_y$ and $l_c=\sqrt{\frac{\hbar}{eB}}$ is the magnetic length. On the harmonic oscillator basis $\phi_n$ with $n=0,1,2,...$, these operators behave as 
\begin{equation}
    a\phi_n=\sqrt{n}\phi_{n-1}, a^{\dagger}\phi_n=\sqrt{n+1}\phi_{n+1}, a^{\dagger}a\phi_n=n\phi_n.
\end{equation}
By replacing $k_x$ and $k_y$ with $a$ and $a^{\dagger}$ in $H_e+H_{e-M}$, we then obtain the Hamiltonian and the wave-function ansatz\cite{novik2005band},

\begin{equation}
    H=
    \begin{pmatrix}
    V_0/2+g M_s m_{s1,z} & -i\sqrt{2}v_f a/l_c & 0 & 0\\
    i\sqrt{2}v_f a^{\dagger}/l_c & V_0/2-g M_s m_{s1,z}  & 0 & 0\\
    0 & 0 & -V_0/2+g M_s m_{s2,z}  & i\sqrt{2}v_f a/l_c\\
    0 & 0 & -i\sqrt{2}v_f a^{\dagger}/l_c & -V_0/2-g M_s m_{s2,z} 
    \end{pmatrix},
\end{equation}
\begin{equation}
    \Psi_N
    =
    exp(-iXy/l_c^2)\begin{pmatrix}
    f_1^N\phi_N \\
    f_2^N\phi_{N+1} \\
    f_3^N\phi_N \\
    f_4^N\phi_{N+1}
    \end{pmatrix},
\end{equation}
where $N=-1,0,1,...$ as $n$ ranges from $0,1,2,...$, $X=-k_yl_c^2$ is the center of motion and $f_i^N$ are the expansion coefficients in the harmonic oscillator basis. From the Hamiltonian with $N=-1$ or $n=0$
\begin{equation}
    H^{n=0}=
    \begin{pmatrix}
    0 & 0& 0 & 0\\
    0 & V_0/2-g M_s m_{s1,z}  & 0 & 0\\
    0 & 0 & 0  & 0\\
    0 & 0 & 0 & -V_0/2-g M_s m_{s2,z} 
    \end{pmatrix},
\end{equation}
\begin{equation}
    \Psi_{n=0}=
    \begin{pmatrix}
    0 \\
    \phi_0 \\
    0 \\
    \phi_0
    \end{pmatrix},
\end{equation}
we can solve the zeroth Landau levels (zLLs) as $\varepsilon_t=V_0/2-gM_sm_{s1,z}$ and $\varepsilon_b=-V_0/2-gM_sm_{s2,z}$ for the top and bottom surface state, respectively. Using $m_{s1,z}=-1$ in AFM1, $m_{s1,z}=1$ in AFM2 and $m_{s2,z}=-m_{s1,z}$, we therefore obtain the energies of zLLs $\varepsilon_{1,\lambda}^0= \lambda g M_s+ \lambda V_0/2 $ for the AFM1 state and $\varepsilon_{2,\lambda}^0= \lambda g M_s-\lambda V_0/2 $ for the AFM2 state, where $\lambda=+$ and $\lambda=-$ corresponds to the zLL on the conduction band bottom and valence band top, repectively.

The higher Landau levels are obtained from the Hamiltonian
\begin{equation}
    H^{n}=
    \begin{pmatrix}
    V_0/2+g M_s m_{s1,z} & \frac{-iv_f\sqrt{2n}}{l_c} & 0 & 0\\
    \frac{iv_f\sqrt{2n}}{l_c} & V_0/2-g M_s m_{s1,z}  & 0 & 0\\
    0 & 0 & -V_0/2+g M_s m_{s2,z} & \frac{iv_f\sqrt{2n}}{l_c}\\
    0 & 0 & \frac{-iv_f\sqrt{2n}}{l_c} & -V_0/2-g M_s m_{s2,z} 
    \end{pmatrix},
\end{equation}
given by $\varepsilon_{\mu,\nu}^n =\mu\sqrt{2v_f^2n/l_c^2+(g M_s)^2} +\nu V_0/2$, where $\mu,\nu=\pm1$, $n=1,2,\cdots$ representing the $n$th LL. All the higher LLs of the two surface states are equivalent for the two AFM configurations, even in the presence of both electric and magnetic fields. 

\subsection{Thin film model} \label{sec:thin_film_model}
\begin{figure}
    \centering
    \includegraphics[width=\textwidth]{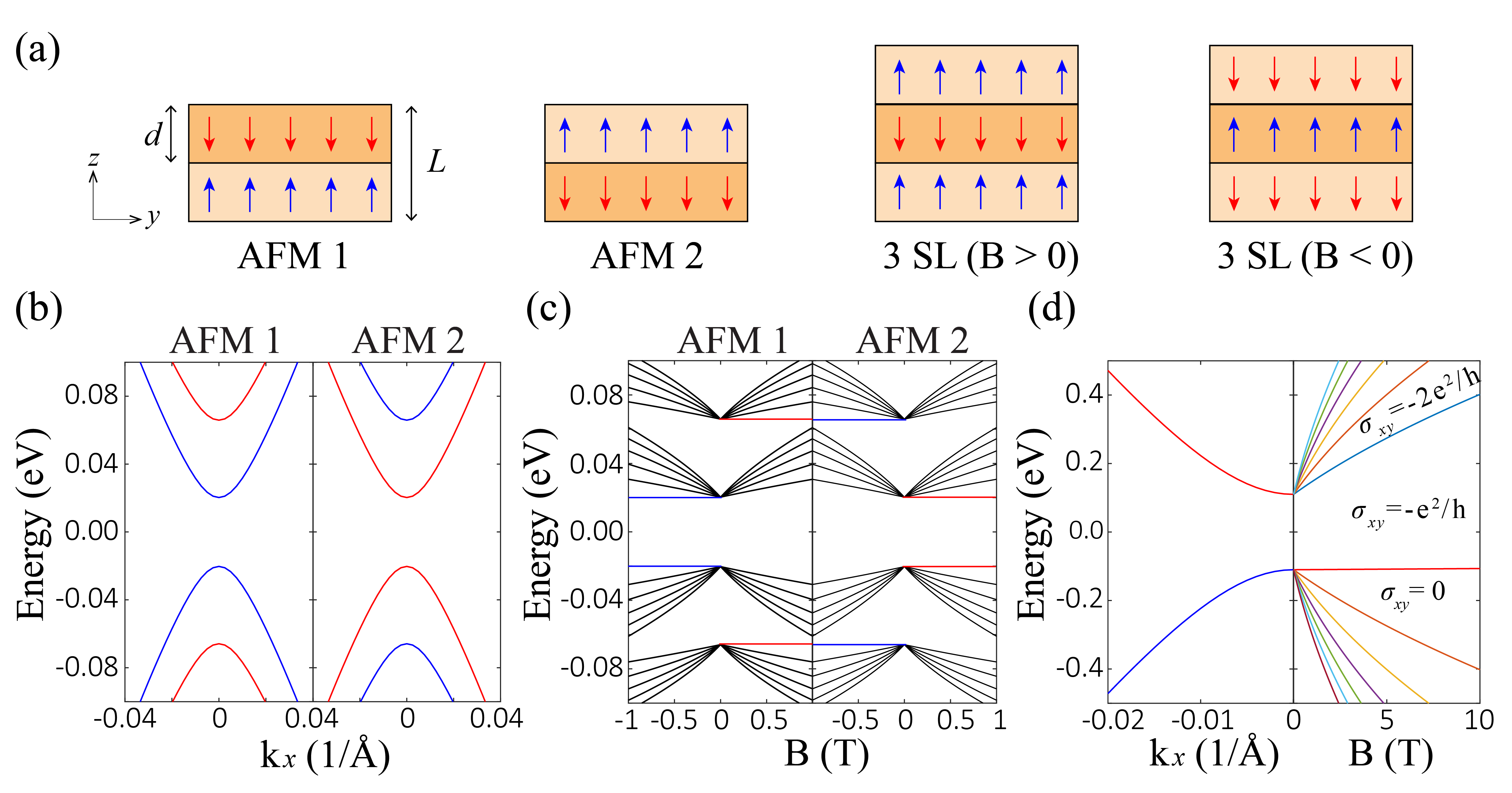}
    \caption{(a) Illustrations of the thin film model for 2 SL and 3 SL MBT films. The samples have a total thickness of $L$ with each septuple layer thickness $d=1.3nm$. The neighboring layers exhibit opposite magnetization to describe AFM order. (b) Electronic energy dispersion for the AFM1 and AFM2 configurations with $g M_s=0.05eV$ and $V_0=0.1eV$. (c) The energy spectrum of Landau levels as a function of magnetic field $B$ for the AFM1 and AFM2 configurations. The zeroth Landau levels are labeled in red and blue, while all the higher Landau levels are labeled in black. (d) Dispersion and Landau levels for 3 SL at $g M_s=0.05eV$ and $V_0=0.1eV$. The Hall conductance is $\sigma_{xy}=-\frac{e^2}{h}$ in the gap. Note that here we fix chemical potential instead of electron density when considering the Hall resistance at different magnetic fields.
    }
    \label{fig:figS2}
\end{figure}

We then considered a MBT thin film based on an effective bulk four-band model \cite{liu2010model}. The intrinsic AFM orders are implemented such that the neighboring layers have opposite exchange coupling $g$. To simplify the case, here we introduce the 2 SL film, with a total thickness $L$ and septuple-layer (SL) thickness $d$ (Fig. \ref{fig:figS2}(a)). The samples with any layers can be calculated with the same method.

The effective four-band Hamiltonian for 2 SL film writes as
\begin{equation}
\begin{gathered}
    H=H_0+H_1+H_2,\\
    H_0=M(\mathbf{k})\tau_z+Bk_z\tau_y+A((k_y+\frac{e}{\hbar}  A_y) \tau_x\sigma_x-(k_x+\frac{e}{\hbar}  A_x) \tau_x\sigma_y),\\
    H_1=-e(Ez-Ed)\sigma_0=(V_0 z/L-V_0/2)\sigma_0,\\
    H_2=
    \begin{cases}
    g M_s m_{s1,z} \sigma_z,& L/2<z<L,\\
    g M_s m_{s2,z} \sigma_z,& 0<z<L/2
    \end{cases},
\end{gathered}
\end{equation}
where the four-band model $H_0$ is written in the basis $|P1_-^+,\uparrow \rangle$, $|P2_+^-,\uparrow \rangle$, $|P1_-^+,\downarrow \rangle$ and $|P2_+^-,\downarrow \rangle$ with Pauli matrices $\tau$ and $\sigma$ in the orbital ($P1, P2$) and spin ($\uparrow, \downarrow$) sub-space, $M(\mathbf{k})=M_0+M_1k_z^2+M_2k_{\parallel}^2$, $E$ is the electric field strength and $V_0$ is the asymmetric potential drop across the whole thin film. 
For the thin film model, the $z$-directional momentum $k_z$ should be replaced by $-i\partial_z$ with open boundary condition $\psi_{k_{\parallel}}(z=0)=\psi_{k_{\parallel}}(z=L)=0$, and the wavefunction is then expanded as
\begin{equation}
    \psi_{k_{\parallel}}(z)=\sum_{n,\lambda}a_{n,\lambda}(k_{\parallel}) \sqrt{\frac{2}{L}} sin(\frac{n\pi}{L}z) |\lambda \rangle,
\end{equation}
where $n=1,2,...,N$, $a_{n,\lambda}(k_{\parallel})$ is the expansion coefficient and $|\lambda \rangle$ is the basis of the four-band model with $\lambda=1,...,4$. Combined with the eigen-equation $H(k_{\parallel})\psi_{k_{\parallel}}(z)=E\psi_{k_{\parallel}}(z)$, we obtain the matrix element of expansion
\begin{equation}
    \langle n,\lambda| H |n',\lambda'\rangle=\frac{2}{L}\int_0^L dz sin(\frac{n\pi}{L}z) H_{\lambda \lambda'} (z) sin(\frac{n'\pi}{L}z).
\end{equation}
 We chose the parameters as $M_0=-0.28eV$, $M_1=6.86eV$\AA$^2$, $M_2=44.5eV$\AA$^2$, $A=3.33eV$\AA, $B=2.26eV$\AA, $N=15$ and $d=1.3nm$ \cite{liu2010model}. 

We examined the effect of $E$ and $B$ on determining the favored AFM configuration by calculating Landau levels following the same method as shown in the last section. The difference is that now the Hamiltonian has $z$ dependence, and thus the eigen-wave function ansatz should take the form \cite{novik2005band}
\begin{equation}
    \Psi_N=
    exp(-iXy/l_c^2)\begin{pmatrix}
    f_1^N(z) \phi_N \\
    f_2^N(z) \phi_{N} \\
    f_3^N(z) \phi_{N+1} \\
    f_4^N(z) \phi_{N+1}
    \end{pmatrix},
\end{equation}
where all the $f_i^N$ are z-dependent functions. 
By solving the Hamiltonian, we plot the dispersion and the energy of LLs as a function of $B$ at $g M_s=0.05eV$ ($g>0$) and $V_0=0.1eV$, as shown in Fig. \ref{fig:figS2}(b) and (c). 
We summarized the energy difference $\Delta\varepsilon=\varepsilon_{AFM 1}-\varepsilon_{AFM 2}$ between two configurations as functions of both $B$ and $V_0$ in Fig. 2(b) in the main text. 

In Fig. \ref{fig:figS2}(d), we show the dispersion and Landau levels for 3-SL MBT film with $g M_s=0.05eV$ ($g>0$) and $V_0=0.1eV$. As the zero-field AH resistance and the ordinary Hall resistance share the same sign for the electron doping in 3-SL MBT film in experiments, this fact is consistent with our Landau level spectrum in Fig. \ref{fig:figS2}(d), in which the zeroth Landau level coincides with the top of the valence bands at small magnetic fields. In our calculation for Fig. \ref{fig:figS2}(d), we choose $g>0$, while the zeroth Landau level will appear at the conduction band bottom if $g<0$. Therefore, by comparing the numerical results with the experiment facts, we can fix the sign of the exchange coupling constant as positive within the definition of our model Hamiltonian, and this allows us to determine the anomalous Hall sign in the next section.

The Hall conductance shown in the main text Fig. 2 is calculated by 
\begin{equation}
 \sigma_{xy}=-i\frac{1}{2\pi} \sum_{filled \ n} \int dk_x dk_y \sum_m \frac{\langle u_n | \partial_x H | u_m \rangle \langle u_m | \partial_y H | u_n \rangle - \langle u_n | \partial_y H | u_m \rangle \langle u_m | \partial_x H | u_n \rangle}{(\varepsilon_n- \varepsilon_m)^2},
\end{equation}
where $\partial_j=\frac{\partial}{\partial k_j}$ ($j=x,y$), $\varepsilon_n$ and $u_n$ are eigenvalues and eigenvectors of the Hamiltonian and $n, m$ are band numbers.

\section{Anomalous Hall Sign} \label{sec:AHsign}

Below we will provide a discussion on the anomalous Hall sign of two AFM configurations in the two-surface-state model. 
From the Landau level calculations in Sec. II.A for the two-surface-state model, we notice that the zeroth LLs only appear for half of the bands and their locations depend on the sign of $g$ and $m_{i,z}$, with $\varepsilon_t=V_0/2-gM_sm_{1,z}$ and $\varepsilon_b=-V_0/2-gM_sm_{2,z}$ for the top and bottom surface states. We first focus on the $g>0$ case. For $B>0$ and $V_0=0$, the magnetization vectors $m_{i,z}$ in odd SL (FM alignment of magnetization between two surfaces) tend to align with the magnetic field, meaning that $m_{1,z}=m_{2,z}=+1$, and thus the zLLs coincide with the energy of valence band top on both top and bottom surfaces with $\varepsilon_t=-g M_s$ and $\varepsilon_b=-g M_s$. In AFM1 with $m_{1,z}=-1$ and $m_{2,z}=+1$, the zLLs are located at the conduction band bottom on the top surface $\varepsilon_t=g M_s$, and valence band top on the bottom surface $\varepsilon_b=-g M_s$. In AFM2, the zLLs are located at the valence band top on the top surface $\varepsilon_t=-g M_s$ 
and conduction band bottom on the bottom surface $\varepsilon_b=g M_s$. 
Meanwhile, we label the AH sign in terms of color red and blue for negative and positive AH signs, respectively, as shown in Fig. \ref{fig:figS1}(b). In odd SL MBT films, the valence bands are assigned with negative AH sign (red) and the conduction bands are assigned with positive AH sign (blue), and thus the zero-field AH resistance should be negative, which is the same as the ordinary Hall resistance (negative for $B>0$), to be consistent with the AH sign determined from Fig. \ref{fig:figS2}(d), as compared with experiments. As the top surface state of AFM1 and the bottom surface state of AFM2 have opposite magnetization compared to the odd SL, the AH signs of their bands switch correspondingly, namely, the valence band is positive (blue) and the conduction band is negative (red), as shown in Fig. \ref{fig:figS1}(b). Furthermore, the position of the zLL of these surface states also change from valence to conduction band as $m_{i,z}$ changes sign, and therefore, all the zLLs are located at the red bands for $g>0$ and $B>0$ (Fig. \ref{fig:figS1}(b)). 

\begin{figure}
    \centering
    \includegraphics[width=\textwidth]{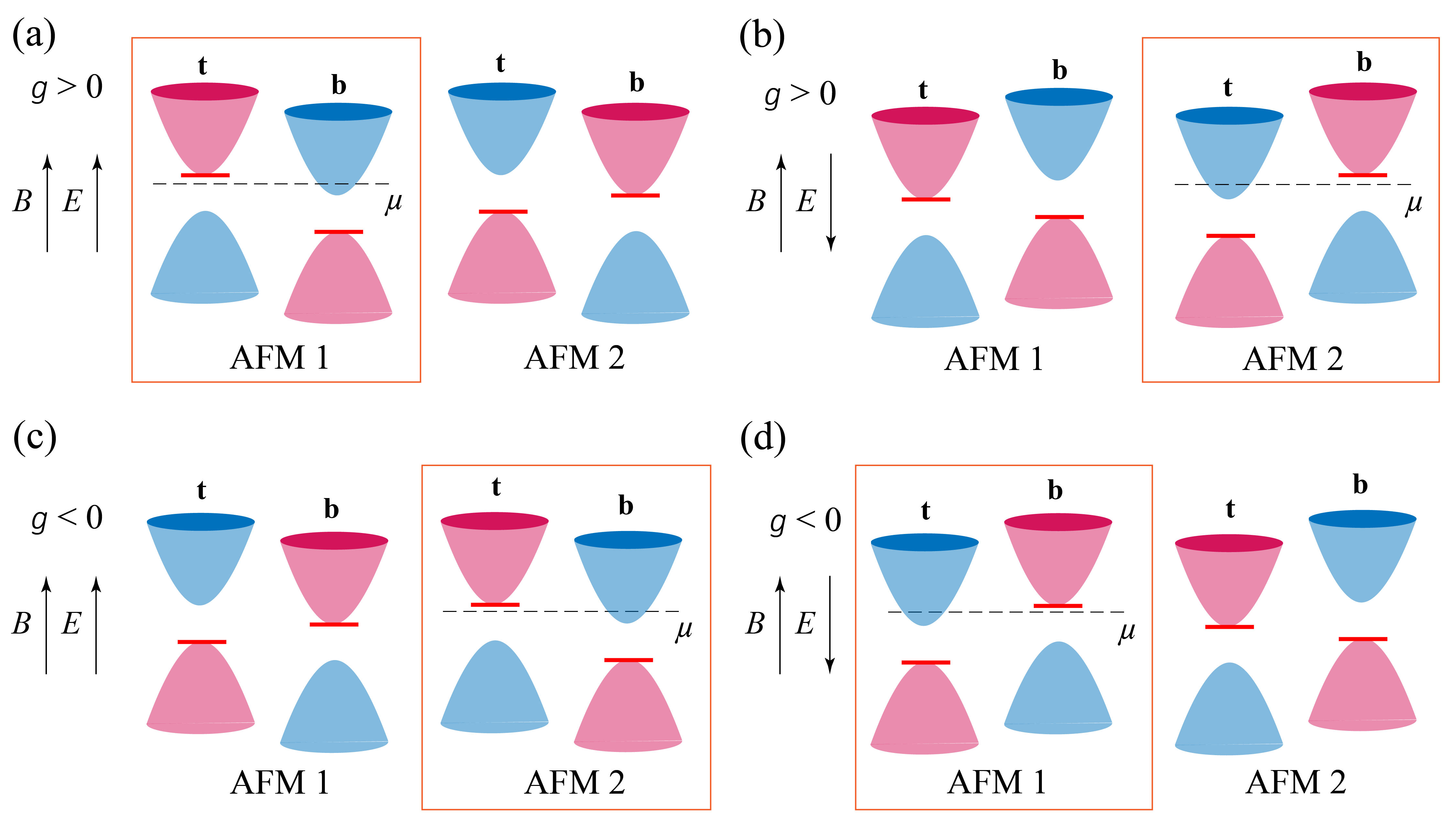}
    \caption{ Illustration of two-surface-state model for AFM1 and AFM2 at (a) $g>0$ and $B>0$ and $E>0$, (b) $g>0$, $B>0$ and $E<0$, (c) $g<0$ and $B>0$ and $E>0$, (d) $g<0$, $B>0$ and $E<0$. The favored AFM configuration in each case is marked by the box.}
    \label{fig:figS3}
\end{figure}

Next we will determine the favored AFM configuration for even SL and its AH sign under the presence of an asymmetric potential. The external electric field splits the energies of the top and bottom surface states, and subsequently differentiates the energies of the occupied zLL in AFM1 and AFM2. If electric field lowers the energy of the bottom surface states but increases the energy of the top surface states ($E>0$ and $V_0>0$), then the energy of the occupied zLL for AFM1 $\varepsilon_b=-V_0/2-g M_s$ decreases, 
while the energy of the occupied zLL for AFM2 $\varepsilon_t=V_0/2-g M_s$ increases. Thus, AFM1 becomes energetically favored for $g>0$, $B>0$ and $E>0$ (Fig. \ref{fig:figS3}(a)). More generally, the energetically favored AFM configuration should have its zLL located at the valence band top of the surface state whose energy is lowered by the electric field.
Since all the zLLs coincide with the red bands, this means that the favored AFM configuration is the one with red valence band on the lower-energy surface state, as shown in Fig. \ref{fig:figS3}(a) and (b). For electron doping in the favored AFM configuration, the electrons will first fill the conduction band of the lower-energy surface states which should have blue color, as the corresponding valence band has red color. As a result, the overall AH sign is positive (blue) in the favored even SL configuration, opposite to the AH sign in odd SL, which is always negative for $g>0$ and $B>0$. Therefore, the even and odd SL have opposite AH signs for the electron doping, irrespective of the electric field direction. 

%For the hole doping, the blue valence band (positive AH sign) of the higher-energy surface state contribute less to AH conductance, so the overall AH sign will be negative, which is the same as the AH sign of odd SL. 
The situation is dramatically changed for hole doping.
For $B>0$ and $E>0$, the Fermi level now crosses the valence band of the top surface in AFM1, so that all the occupied states contribute a negative AH conductance, which is the same sign of the AH conductance in odd SL. Similar situation occurs for $B>0$ and $E<0$. 
The odd SL and even SL both exhibit negative (positive) AH sign under positive (negative) magnetic field, again regardless of the electric field direction. This is consistent with the experiment by \emph{Gao et al.} \cite{gao2021layer}, where the 5-SL MBT exhibits negative AH at both electron and hole doping, while the 6-SL MBT displays negative and positive AH resistance at hole and electron doping, respectively.  For an magnetic field of $1T$ and electric field of $0.2V/nm$ ($V_0 \approx 1eV$ for 6 SL MBT film) corresponding to the values in \emph{Gao et al.}, the scale of the energy difference from our calculation is around $\Delta\varepsilon/n=0.1meV$ (using $\Delta\varepsilon=1eV/\mu m^2$ and $n=1\times 10^{12}cm^{-2}$), which is about 1K at the temperature scale and is at the same order as the temperature  (around 2K) in experiments by \emph{Gao et al.}.

\begin{figure*}
    \centering
    \includegraphics[width=\textwidth]{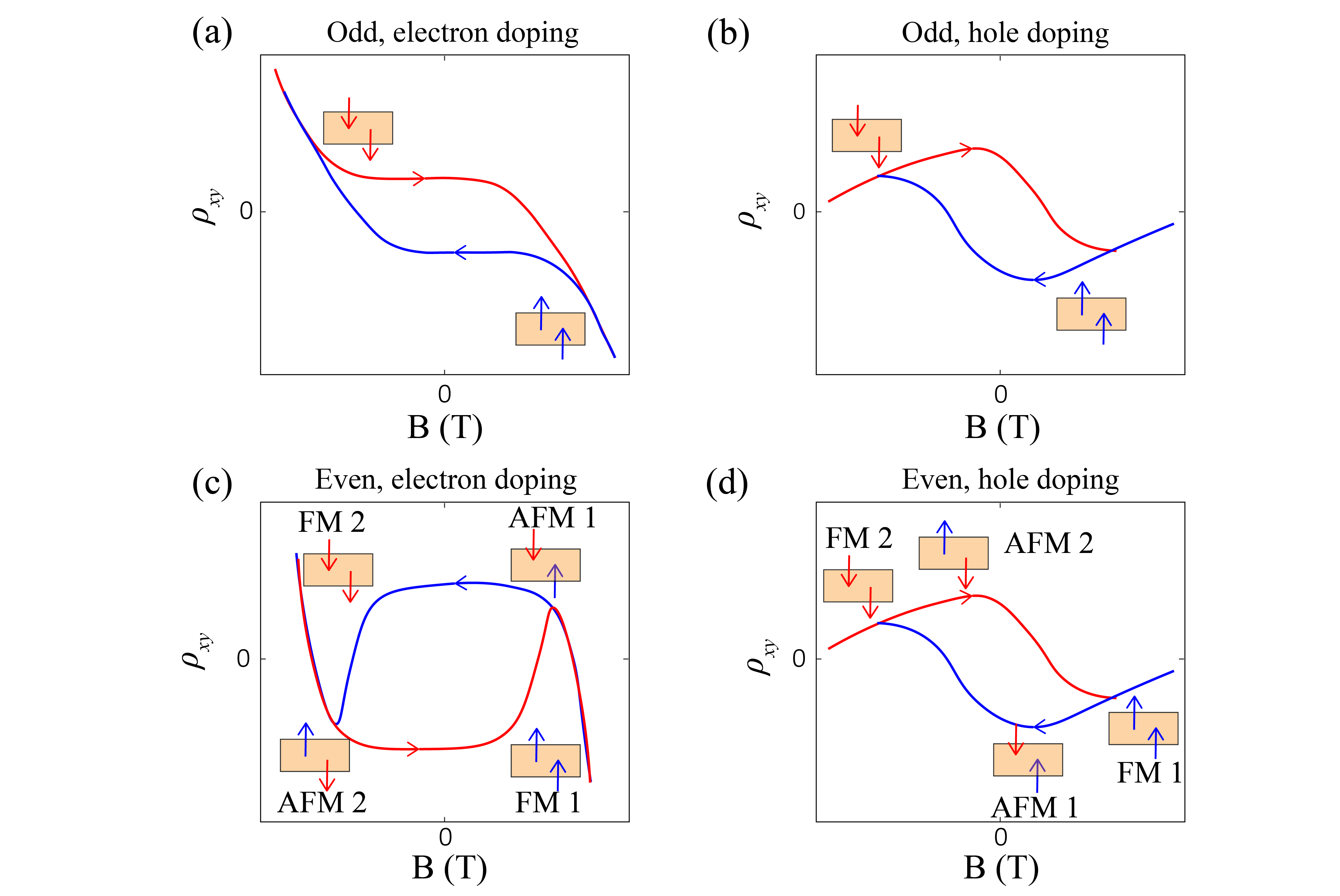}
    \caption{Predictions of AH hysteresis loops for odd and even SL MBT films at electron and hole doping for $g>0$. (a) Odd SL at electron doping. The AH sign changes from negative to positive when $B$ sweeps from positive to negative. (b) Even SL at electron doping. The favored AFM configuration has a positive AH sign, so it changes from negative to positive as the system transits from the FM state into AFM state when $B$ decreases from a positive value. Notably, there exists a sign change originating from the transition between two AFM states before the spin-flop transition. (c) Odd SL at hole doping, at which the ordinary Hall has positive slopes. (d) Even SL at hole doping. It shows a similar pattern as the odd SL since they have the same AH sign.}
    \label{fig:figS4}
\end{figure*}

Based on the above analysis, we sketch the hysteresis loops for odd and even SL at electron and hole dopings in Fig. \ref{fig:figS4}, and we notice that only the even SL at electron doping shows different zero-field AH sign compared to the other cases. For odd SL at electron doping, the ordinary Hall and the AH resistance share the same sign (Fig. \ref{fig:figS4}(a)); while at hole doping, the ordinary Hall resistance has opposite sign as the AH resistance (Fig. \ref{fig:figS4}(b)) and as a result there is a sign change when the magnetic field reduces from a large positive value toward zero. For even SL at electron doping, the AH of FM state and the ordinary Hall have the same sign, and the sign changes happen at a transition from FM state to AFM state and another transition between two AFM states (Fig. \ref{fig:figS4}(c)), which has been discussed in details in the main text. At hole doping, the FM state and AFM state of even SL share the same AH sign, while the ordinary Hall switches sign, and therefore when the magnetic field is reduced from a large positive value toward negative, the Hall resistance changes sign first from ordinary Hall to FM state, followed by another sign change caused by transition between two AFM states (Fig. \ref{fig:figS4}(d)).

On the other hand, for the $g<0$ case, the location of zLLs in odd SL changes to the conduction band bottom on both surfaces when $g$ changes sign (for $B>0$ in Fig. \ref{fig:figS1}(c)). Meanwhile, the AH signs of the bands change as well, since the AH signs depend on the sign of $g m_{i,z}$. Therefore, in odd SL, the valence bands have positive AH sign and the conduction bands have negative sign, and thus the zero-field AH resistance should be positive when the Fermi energy is around the gap for $g<0$. We notice this AH sign in this case is opposite to the ordinary Hall sign at the electron doping, so it is inconsistent with the experimental observations. But we will still discuss the possible scenario for the even SL case when $g<0$. 

For AFM1 of the even SL case, the zLLs are located at the valence band top on the top surface $\varepsilon_t=g M_s$ ($M_s$ is always assumed to be positive), and the conduction band bottom on the bottom surface $\varepsilon_b=-g M_s$. For AFM2 of the even SL case, they are located at the conduction band bottom on the top surface $\varepsilon_t=-g M_s$ and valence band top on the bottom surface $\varepsilon_b=g M_s$.  
Thus, no matter the sign of $g$, the locations of the zLLs always coincide with the band energy extreme of the red bands in Fig. \ref{fig:figS3}.
Similar to the discussion above for the $g>0$ case, we find the energetically favored state for the $g<0$ case is AFM2 when $B$ and $E$ are parallel, and AFM1 when $B$ and $E$ are anti-parallel, in Fig. \ref{fig:figS3}(c) and (d). At electron doping, we notice that the Fermi energy will always cross the blue conduction band of the energetically favored magnetic configuration, no matter this configuration is AFM2 (Fig. \ref{fig:figS3}(c)) or AFM1 (Fig. \ref{fig:figS3}(d)). Consequently, the AH sign is always positive at electron doping, determined by the blue conduction bands, in both Fig. \ref{fig:figS3}(c) and (d). Therefore, for $g<0$, the even and odd SL show the same positive AH sign at electron doping; while at hole doping, the odd SL still has positive AH sign but the even SL has negative AH sign.

The sign of the exchange coupling $g$ can be determined from comparing the signs of the ordinary Hall and AH resistance in odd SL to those in experiments, as discussed in Sec. II.B For $g>0$, the ordinary Hall and AH share the same sign, which is consistent with experimental measurements, while for $g<0$, they have opposite signs for odd SLs. Therefore, we can fix the sign of exchange coupling as positive (within our definition of the model). Our above analysis further suggests that the odd and even SL films show opposite AH signs at electron doping and the same sign at hole doping for $g>0$, which is also consistent with experimental observations.

It was predicted that even SL films are axion insulators while odd SL films are QAH insulators when the MBT films are insulating \cite{otrokov2019unique,li2019intrisic,zhang2019topological}. We notice that the experiments for insulating MBT films are controversial \cite{chang2023colloquium} and this simple picture for insulating MBT films has not been experimentally confirmed, as the real samples are quite dirty and complicated. For example, electron-hole paddles, Mn doping and film thickness fluctuations can exist in MBT films and can have a strong influence on the transport property for the “insulating” regime. Our intrinsic mechanism for the even-odd effects in the electron and hole doping regions of MBT films can be naturally connected to the even-odd effect in the insulating regime. For even SL MBT films, the AH signs are predicted to be opposite for electron and hole doping, so that the AH conductance is naturally zero when the Fermi energy is tuned to the insulating regime. For odd SL MBT films, both electron and hole doping give rise to the same AH sign as the insulating case. Due to the large free carriers in metallic MBT films, we expect the influence of sample inhomogeneity and Mn doping becomes weaker so that the even-odd effect is more robust.

\section{Multi-step Spin-flop transition in thicker even SL MBT films}
For even SL MBT films thicker than 2 SL, multiple intermediate magnetic configurations can exist and have been studied in literature \cite{yang2021odd,sass2020robust}. As a result, a series of subsequent spin-flop transitions between different states may occur, as evidenced in Fig. 2d-h in Ref. \cite{yang2021odd}, where multi-step spin-flop transitions were observed. Previous works also find that the surface spin-flop transitions happen at a smaller magnetic field than the bulk spin-flop transitions in even MBT films as the magnetic field increases (Fig. 3c in Ref. \cite{bac2022topological}, Fig. 4d in Ref. \cite{sass2020robust} and Fig. 2h in Ref. \cite{yang2021odd}). The transition between AFM configurations discussed in this paper occurs at an even smaller magnetic field than that of the surface spin-flop transition. We notice that the hysteresis loop of Hall resistance $R_{xy}$ can exist in even SL MBT films with the coercive magnetic field smaller than that for all different types of spin-flop transition Fig. 4e in Ref. \cite{chen2019intrinsic}, Fig. 2b in Ref. \cite{ovchinnikov2021intertwined} and Fig. 1e in Ref. \cite{yang2021odd}. Previous experiments attributed this observation as the thickness-independent surface related magnetization \cite{yang2021odd}, defects and disorders in the samples, or substrate-induced top-bottom surface asymmetry \cite{zhao2021even,chen2019intrinsic}. Our theoretical studies provide an intrinsic mechanism of the transition between two AFM states to understand this hysteresis loop in even SL MBT films. 

\begin{figure}
    \centering
    \includegraphics[width=\textwidth]{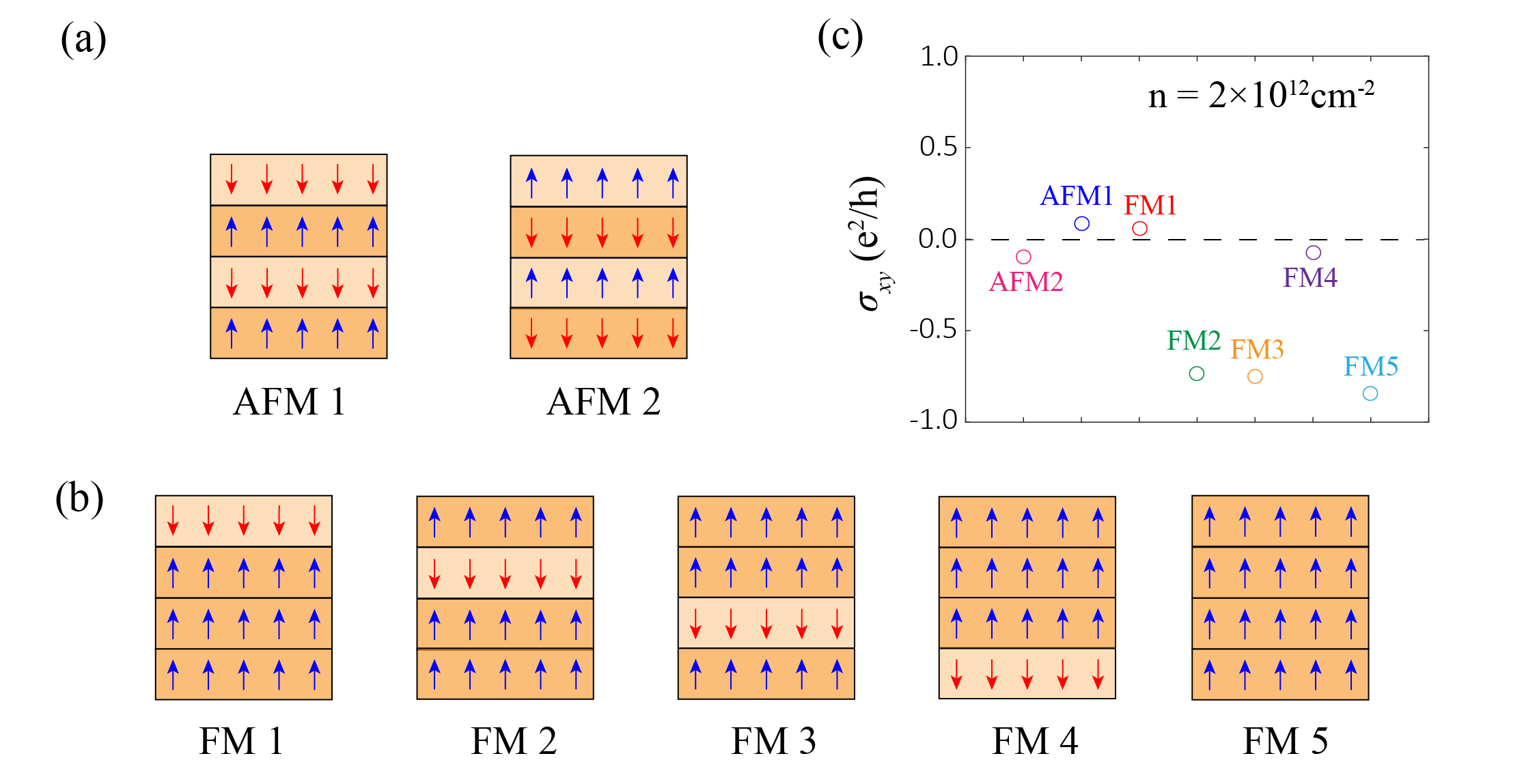}
    \caption{(a) Two AFM states of 4 SL MBT film. (b) Possible FM configurations of 4 SL MBT film for $B>0$. (c) Calculated Hall conductance for each state at $V_0=0.03eV$ and $n=2 \times 10^{12}cm^{-2}$. }
    \label{fig:figS8}
\end{figure}

Below we take a 4 SL MBT film as an example and provide an understanding of the hysteresis loops and multi-step spin-flop transitions in multilayer MBT films.
First, near zero magnetic field, the two AFM states with opposite magnetization in the neighboring layers are the ground state (Fig. \ref{fig:figS8}a), as studied in previous works \cite{ovchinnikov2021intertwined,zhang2019topological}. Therefore, the first transition is between these two AFM states, determined by the direction of external electric and magnetic fields. We calculated the Hall conductance of AFM 1 and AFM 2 for 4 SL in the thin film model at $V_0=0.03eV$ and $n=2\times 10^{12}cm^{-2}$ (Fig. \ref{fig:figS8}c). When $B>0$ ($B<0$) and $V_0>0$, the favored AFM state is AFM 1 (AFM 2) as discussed in the main text. Thus, as the magnetic field sweeps from negative to positive, the sign of the Hall conductance changes from negative to positive, corresponding to the transition from AFM 2 to AFM 1.

Second, as the magnetic field becomes larger, other possible ferromagnetic (FM) configurations arise, as shown in Fig. \ref{fig:figS8}b for cases with $B>0$, where the overall magnetization is aligned with the external magnetic field. The calculated Hall conductance of these FM states are also shown in Fig. \ref{fig:figS8}c. As mentioned above, the surface spin-flop transition occurs before the bulk spin-flop transitions, so the top surface layer in AFM 1 will first flip spins as the magnetic field increases. As a result, the first spin-flop transition most likely corresponds to the transition between AFM 1 and FM 3, during which the sign of the Hall conductance changes from positive to negative. In real materials, the magnetization is canted for these intermediate magnetic states, but here we omit the canting effect since the sign of Hall conductance only depends on the out-of-plane magnetization. With further increasing magnetic fields, the FM state will transition from FM3 to FM5, leading to the further increasing of AH conductance. This gives rise to the multi-step spin-flop transitions observed in experiments. 

\section{Influence of higher Landau levels}
In experiments, the first (or even higher) Landau levels (LLs) can be occupied. We will discuss below that including these higher LLs will not change our mechanism. We will further derive the constraint of the valid parameter regimes for our mechanism to work. 

\begin{figure}
    \centering
    \includegraphics[width=\textwidth]{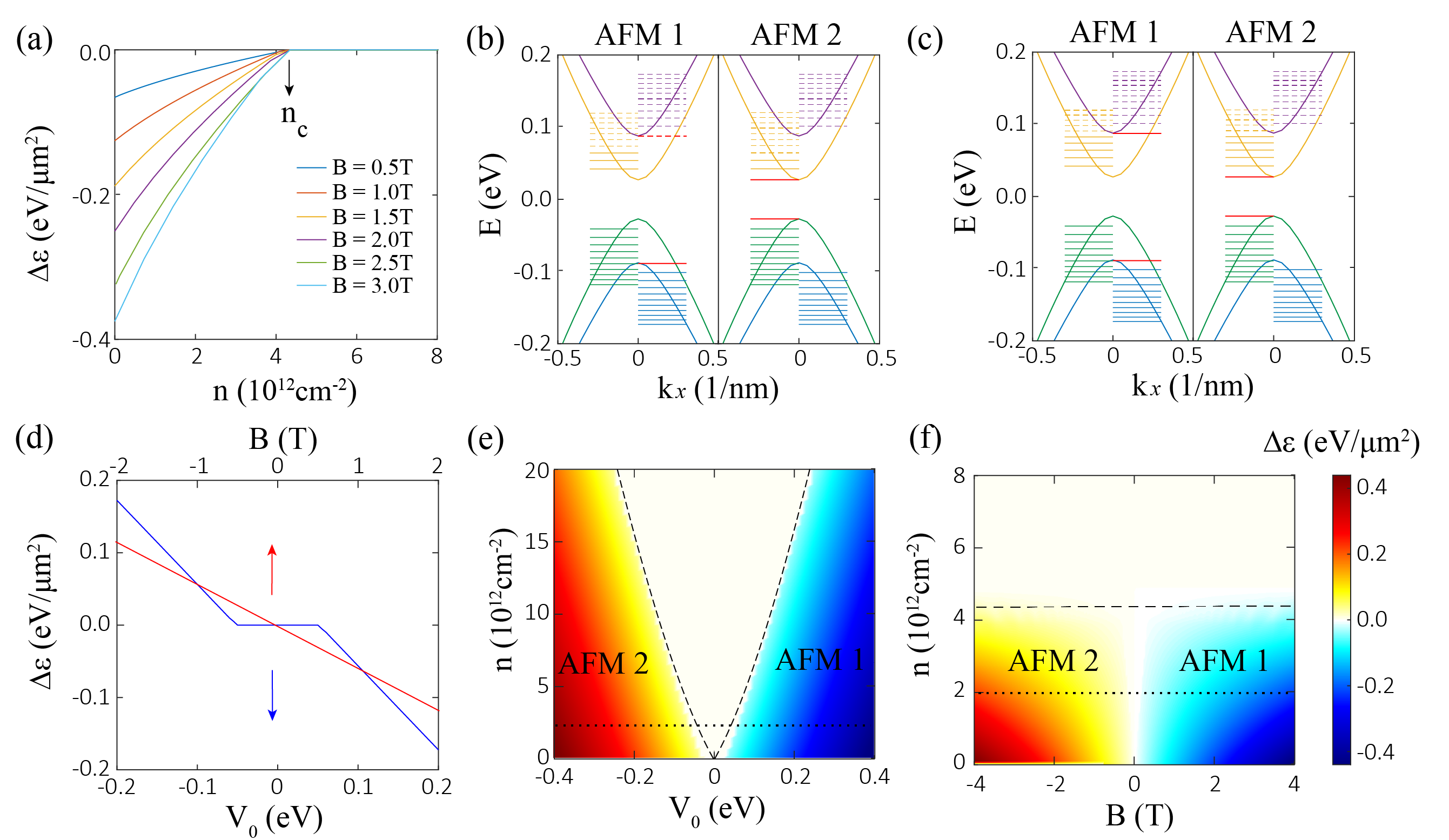}
    \caption{(a) Energy difference between AFM 1 and AFM 2 as a function of carrier density $n$ with different magnetic fields at $V_0=0.1eV$. (b) Dispersion and Landau Levels for $V_0=0.1eV$ and $B=2 T$. Solid lines are filled LLs and dashed lines are empty LLs. The lowest three LLs in the conduction band are filled. (c) Case when the zeroth LL in the conduction band of AFM 1 is filled at an higher electron density. (d) Energy difference between AFM 1 and AFM 2 as a function of $V_0$ (blue curve) at $B=1 T$ and as a function of $B$ (red curve) at $V_0=0.1eV$. The carrier density is fixed at $n=2\times 10^{12}cm^{-2}$.  (e) Energy difference as a function of carrier density $n$ and asymmetric potential $V_0$ at $B=1 T$. (f) Energy difference as a function of $n$ and $B$ at $V_0=0.1eV$.}
    \label{fig:figS11}
\end{figure}

To investigate the situation where the higher Landau levels are occupied, we calculate the energy difference between AFM1 and AFM2 $\Delta\varepsilon$ as a function of the carrier density $n$ at $V_0=0.1eV$ (Fig. \ref{fig:figS11}a). At $n=0$, the energy difference only comes from the occupied zLLs at the valence bands in two AFM states, and is shown in Fig. \ref{fig:figS11}b or c, in which the zeroth LL of valence band in AFM1 has a lower energy than that in AFM2 for our parameter set.  As the carrier density of conduction bands increases, we notice that the absolute value of the energy difference decreases until it reaches zero in Fig. \ref{fig:figS11}a. This is because the zeroth LL of conduction band in AFM2 has lower energy compared to that in AFM1. Due to the asymmetric potential $V_0$ induced by external electric field, we find that for AFM1, the first LL of the bottom surface has the lower energy than the zeroth LL of the top surface (Fig. \ref{fig:figS11}b). Consequently, for the AFM2 configuration, the electrons first fill the zeroth LL of the conduction band while for the AFM1, they fill the first LL of conduction bands whose energy is higher than the zeroth LL of AFM 2, and thus reduce the total energy difference (Fig. \ref{fig:figS11}b, where the solid lines are filled LLs and dashed lines are unfilled LLs). When the carrier density further increases and fills the zeroth LL of the conduction band at top surface in AFM1, as shown in Fig. \ref{fig:figS11}c, the combined energies of the zeroth LLs in two AFM states become the same and the energy difference goes to zero. Thus, from this discussion and the plot in Fig. \ref{fig:figS11}a, we find that for a fixed external electric field (described by $V_0$) and magnetic field $B$, there is a critical electron density $n_c$, below which there is an energy difference between AFM1 and AFM2 so that a hysteresis loop can occur, while above which there is no energy difference between AFM1 and AFM2. This critical electron density is determined by the condition when the zeroth LLs of conduction bands are fully occupied for both the AFM1 and AFM2 configurations.     

This also implies that for a finite carrier density, there is a minimum $V_0$ for the energy difference between AFM1 and AFM2 to be nonzero, as shown in Fig. \ref{fig:figS11}d, where $\Delta\varepsilon$ is plotted as a function of $V_0$ (blue line) for a fixed magnetic field $B=1T$ and carrier density $n=2\times 10^{12}cm^{-2}$. As discussed above, the energy difference between two AFM states comes from the different occupations of the zeroth LLs in two AFM configurations, which can only be achieved when a minimal asymmetric potential is introduced.  This minimal value of $V_0$ increases with the carrier density, as shown in Fig. \ref{fig:figS11}e, where the dashed line labels the critical values for $n$ and $V_0$ and the dotted line corresponds to the blue curve in Fig. \ref{fig:figS11}d.

We also show the energy difference as a function of $B$ in Fig. \ref{fig:figS11}d with $V_0=0.1eV$ and $n=2\times 10^{12}cm^{-2}$, where there is no such constraint for the magnetic field, and a small $B$ is sufficient to induce a difference in the energy between two AFM states. In Fig. \ref{fig:figS11}f, we plotted the energy difference as a function of $n$ and $B$, where the dashed line indicates the critical carrier density for a fixed asymmetric potential $V_0=0.1eV$ and the dotted line corresponds to the red line in Fig. \ref{fig:figS11}d.

\section{Disorder Effect} 
In experiments, magnetic disorders, particularly magnetic domains, must be present. Magnetic disorders will have a strong influence of AH effect. Firstly, magnetic disorders are expected to reduce the AH conductivity if the system is in the metallic regime, as demonstrated in early studies \cite{onoda2006intrinsic}. Moreover, magnetic disorders can also lead to extrinsic mechanisms of AH conductance such as skew scattering, as studied in 
Ref. \cite{zhang2019topological,chen2019intrinsic}. The competition and interplay between intrinsic and extrinsic mechanisms of AH effect in real materials are a complicated topic \cite{nagaosa2010anomalous}, which is beyond the scope of the current manuscript. We notice that in moderately dirty regime of disorders, the intrinsic AH contribution can become the dominant mechanism over extrinsic skew scattering for the AH effect based on the early theoretical studies \cite{nagaosa2010anomalous,onoda2006intrinsic}. As both disorder strength and spin-orbit coupling are strong, we anticipate the MBT films are likely in the regime where the intrinsic mechanism is dominant.

\section{Orbital magnetization in MBT films}

In this section, we look into the orbital magnetization and axion dynamics in MBT films. As mentioned in the main text, the orbital magnetization contains two parts $m_{total}=m_{tri}+m_{topo}$, a trivial part and a topological part,
\begin{equation}\label{eq:mtrivial}
    \begin{split}
        m_{tri}&=-\frac{ie}{2\hbar} \langle \nabla_k u| \times [H(k)-\epsilon(k)]| \nabla_k u\rangle \\
    &=-\frac{ie}{2\hbar} \sum_{u'} \frac{\langle u|\partial_x H| u'\rangle \langle u'|\partial_y H| u\rangle-\langle u|\partial_y H| u'\rangle \langle u'|\partial_x H| u\rangle}{\epsilon'-\epsilon} , 
    \end{split}
\end{equation}
\begin{equation}\label{eq:mtopo}
    \begin{split}
        m_{topo}&=-\frac{ie}{2\hbar} \langle \nabla_k u| \times 2 [\epsilon(k)-\mu]|\nabla_k u \rangle \\
    &=-\frac{ie}{\hbar} [\epsilon(k)-\mu] \sum_{u'} \frac{\langle u|\partial_x H| u'\rangle \langle u'|\partial_y H| u\rangle-\langle u|\partial_y H| u'\rangle \langle u'|\partial_x H| u\rangle}{(\epsilon'-\epsilon)^2},
    \end{split}
\end{equation} 
where $\epsilon$ and $|u\rangle$ are the eigenvalues and eigenstates of the Hamiltonian $H_e+H_{e-M}$
For the two-surface-state model, we can derive the orbital magnetic moments analytically following Ref. \cite{xiao2007valley}. Let us consider only the top surface state $H_t=v_f (k_x \sigma_y - k_y \sigma_x)+g M_s m_{1,z} \sigma_z + V_0/2$. The two eigenvalues are $\epsilon_1=V_0/2-A$ and $\epsilon_2=V_0/2+A$ with $A=\sqrt{(gM_s)^2+v_f^2 k^2}$ and the corresponding eigenstates are 
\begin{equation}
\begin{gathered}
u_1=\frac{1}{\sqrt{v_f^2 k^2+(gM_sm_{1,z}-A)^2}} \begin{pmatrix}
 gM_sm_{1,z}-A\\
ik_+v_f
\end{pmatrix},\\
u_2=\frac{1}{\sqrt{v_f^2 k^2+(gM_sm_{1,z}+A)^2}} \begin{pmatrix}
 gM_sm_{1,z}+A\\
ik_+v_f
\end{pmatrix},
\end{gathered}
\end{equation}
with $k_{\pm}=k_x \pm ik_y$. We can then derive
\begin{equation}
    \langle u_1|\partial_x H| u_2\rangle \langle u_2|\partial_y H| u_1\rangle-\langle u_1|\partial_y H| u_2\rangle \langle u_2|\partial_x H| u_1\rangle = -i\frac{8}{N^2}v_f^4 k_+k_- gM_sm_{1,z}A,
\end{equation}
where $N=\sqrt{(v_f^2k^2+(gM_sm_{1,z}-A)^2)(v_f^2k^2+(gM_sm_{1,z}+A)^2)}=\sqrt{4v_f^2k^2(v_f^2k^2+(gM_s)^2)}$.
Using Eq. (\ref{eq:mtrivial}) and (\ref{eq:mtopo}), we obtain the orbital magnetic moments of the valence band 
\begin{equation}
    m_{tri,top}=-\frac{ev_f^2gM_sm_{1,z}}{2\hbar((gM_s)^2+v_f^2k^2)},
\end{equation}
\begin{equation}
    m_{topo,top}=\frac{ev_f^2gM_sm_{1,z}}{2\hbar((gM_s)^2+v_f^2k^2)^{3/2}}(\sqrt{(gM_s)^2+v_f^2k^2}+\mu-V_0/2)
\end{equation}
for the top surface states. 
Similarly, we can get the analytical results for the bottom surface state $H_b=-v_f (k_x \sigma_y - k_y \sigma_x)+g M_s m_{2,z} \sigma_z - V_0/2$ as
\begin{equation}
    m_{tri,bot}=-\frac{ev_f^2gM_sm_{2,z}}{2\hbar((gM_s)^2+v_f^2k^2)},
\end{equation}
\begin{equation}
    m_{topo,bot}=\frac{ev_f^2gM_sm_{2,z}}{2\hbar((gM_s)^2+v_f^2k^2)^{3/2}}(\sqrt{(gM_s)^2+v_f^2k^2}+\mu+V_0/2).
\end{equation}
Therefore, the sum of contributions from the top and bottom surface states is
\begin{equation}\label{eq:mtrivial2}
    m_{tri}=-\frac{ev_f^2gM_s}{2\hbar((gM_s)^2+v_f^2k^2)}(m_{1,z}+m_{2,z}),
\end{equation}
\begin{equation}\label{eq:mtopo2}
m_{topo}=\frac{ev_f^2gM_s}{2\hbar((gM_s)^2+v_f^2k^2)^{3/2}}[(m_{1,z}+m_{2,z})(\sqrt{(gM_s)^2+v_f^2k^2}+\mu)+(m_{2,z}-m_{1,z})V_0/2].
\end{equation}
As a result, for odd SL ($m_{1,z}=m_{2,z}=+1$) at $\mu=0$, there is $m_{tri}=-\frac{ev_f^2gM_s}{\hbar((gM_s)^2+v_f^2k^2)}$ and $m_{topo}=\frac{ev_f^2gM_s}{\hbar((gM_s)^2+v_f^2k^2)}$ which cancel out exactly with each other and lead to zero total orbital magnetization. This is consistent with the result in Ref. \cite{xiao2007valley}. For even SL ($m_{1,z}=-m_{2,z}$), we can see that the trivial part is zero while the topological part is $m_{topo}=\frac{ev_f^2gM_s m_{2,z}V_0}{2\hbar((gM_s)^2+v_f^2k^2)^{3/2}}$, with $m_{2,z}=1$ for AFM1 and $m_{2,z}=-1$ for AFM2. 
We emphasize that these results are only valid when there is no coupling between two surface states, while in the numerical calculations of the thin-film model in the main text, a small trivial part of orbital magnetic moment can still be induced due to the hybridization between two surface states, as discussed below.

\begin{figure}
    \centering
    \includegraphics[width=\textwidth]{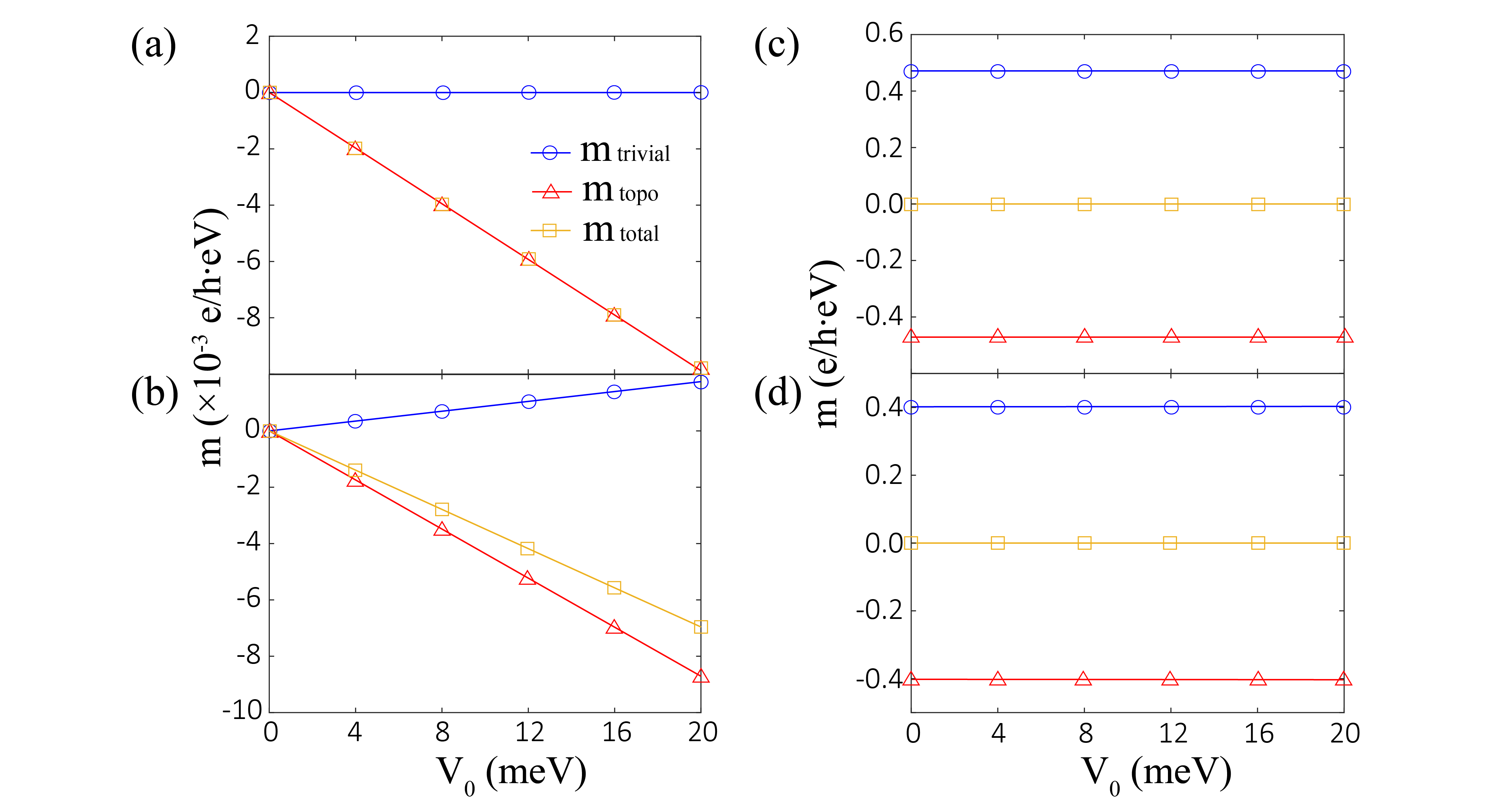}
    \caption{Orbital magnetic moment density $m$ as a function of asymmetric potential $V_0$ in the two-surface-state model at $\mu=0eV$ and $g M_s=0.1eV$ ($g>0$). (a) AFM1 with $t=0eV$. The trivial part remains zero while the topological part increases linearly with $V_0$. (b) AFM1 with $t=0.1eV$. Both trivial and topological part varies linearly with $V_0$ but with opposite signs of slopes. (c) Odd SL with $t=0eV$. Both trivial and topological part are constant and their sum $m_{total}$ is zero. (d) Odd SL with $t=0.1eV$. The trivial and topological part still present zero slopes and cancel out with each other exactly.}
    \label{fig:figS5}
\end{figure}

\begin{figure}
    \centering
    \includegraphics[width=\textwidth]{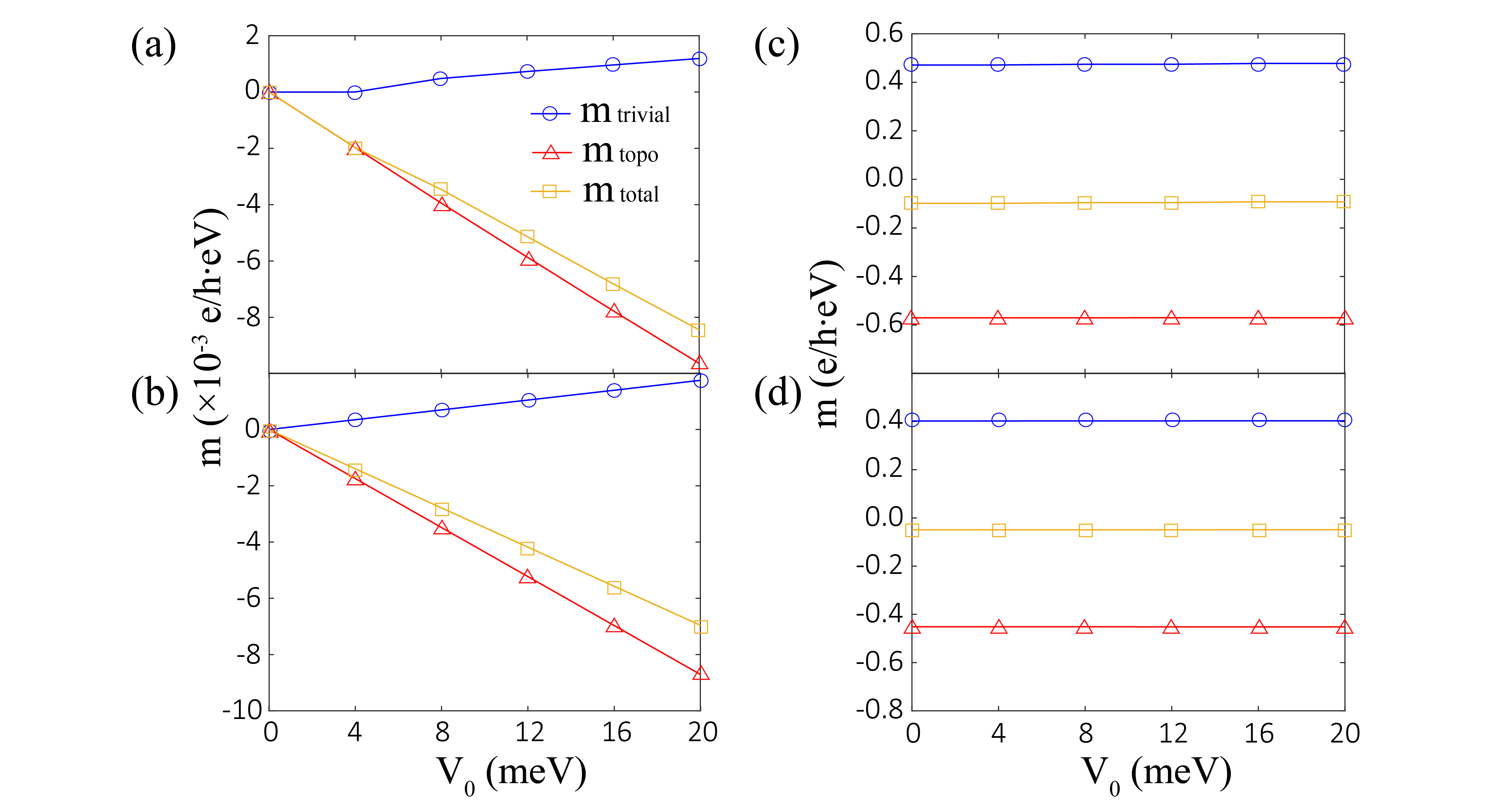}
    \caption{Orbital magnetic moment density $m$ as a function of asymmetric potential $V_0$ in the two-surface-state model at $\mu=0.1eV$ and $g M_s=0.1eV$. (a) AFM1 with $t=0eV$. (b) AFM1 with $t=0.1eV$. (c) Odd SL with $t=0eV$. The total magnetic moment now has nonzero value. (d) Odd SL with $t=0.1eV$.}
    \label{fig:figS6}
\end{figure}

\begin{figure}
    \centering
    \includegraphics[width=\textwidth]{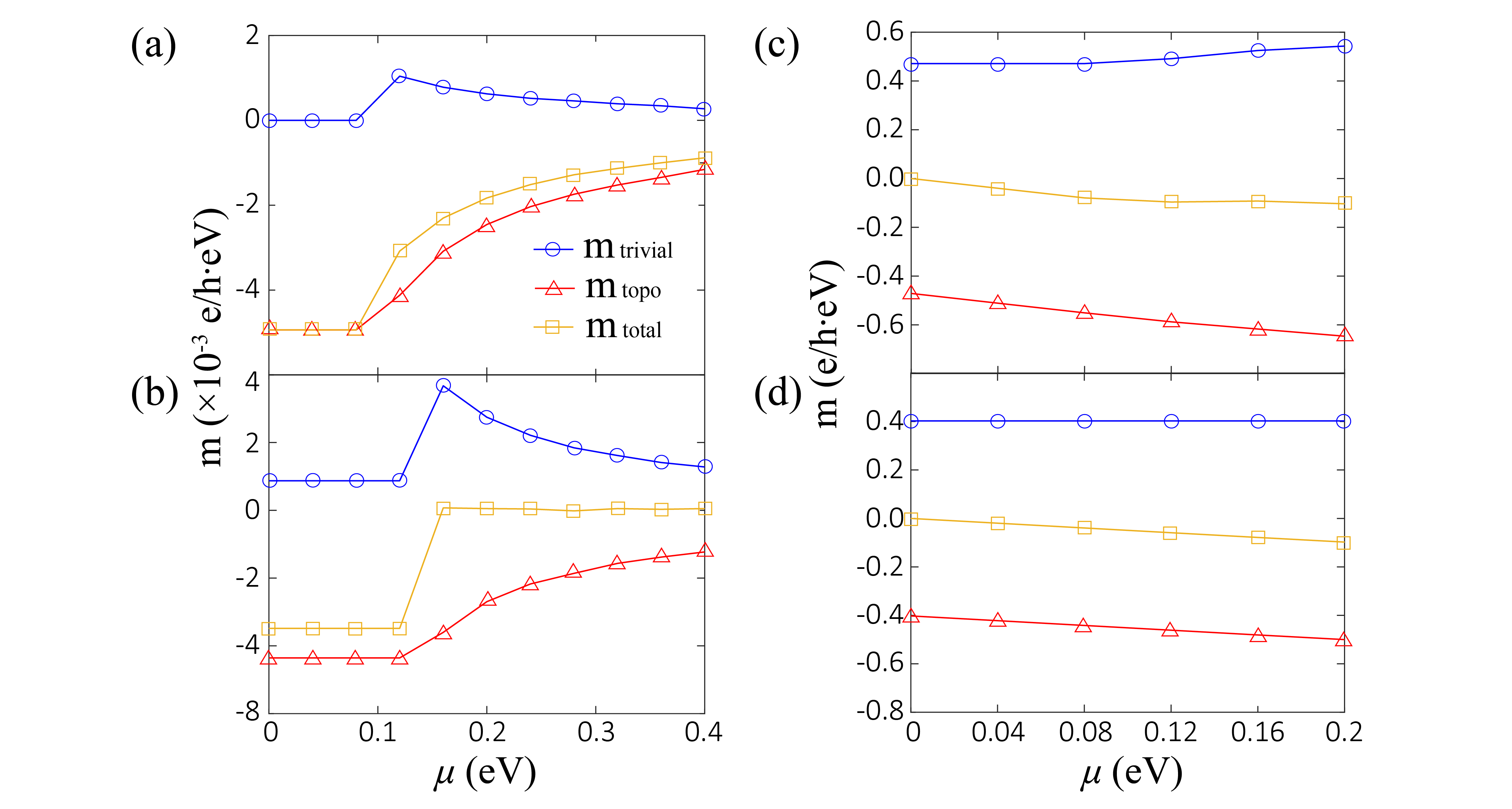}
    \caption{Orbital magnetic moment density $m$ as a function of chemical potential $\mu$ in the two-surface-state model at $V_0=10meV$ and $g M_s=0.1eV$. (a) AFM1 with $t=0eV$. (b) AFM1 with $t=0.1eV$. (c) Odd SL with $t=0eV$. (d) Odd SL with $t=0.1eV$.}
    \label{fig:figS7}
\end{figure}

By evaluating the orbital magnetization, we can also determine the favored AFM state, which is quantitatively equivalent to the zeroth LLs perspective. As a simple example, the energies of the occupied zeroth LLs are $\varepsilon_{1,-}^0=-gM_s-V_0/2$ for AFM 1 and $\varepsilon_{2,-}^0=-gM_s+V_0/2$ for AFM 2, as calculated in Appendix Sec. II. A, for the two-surface-state model. Therefore, the energy difference (per area) between them in the insulating region is $\Delta E=\frac{1}{2\pi l_c^2}(\varepsilon_{1,-}^0-\varepsilon_{2,-}^0)=-\frac{e}{h} BV_0$, where $l_c=\sqrt{\frac{\hbar}{eB}}$ is the magnetic length. 
On the other hand, we show that the energy difference (per volume) between the two AFM states is $\Delta\varepsilon=-2M_0 B$ in the manuscript from the perspective of orbital magnetization. For thick even MBT films (ideal axion insulator), there is $M_0=\alpha E=\frac{e^2}{2h} E$ in the insulating region. Thus, the energy difference (per area) is $\Delta E=\Delta \varepsilon L=-\frac{e^2}{h} ELB=-\frac{e}{h}BV_0$, which is the same result from the zeroth LL approach.

The hybridization between two surface states can be taken into account by the Hamiltonian $H_t=t\tau_x$ in addition to the Hamiltonian (S1) and numerically calculate orbital magnetic moment density in the two-surface-state model. Fig. \ref{fig:figS5} and Fig. \ref{fig:figS6} plot the two parts of the orbital magnetic moment and their sum as a function of $V_0$ at $\mu=0eV$ and $\mu=0.1eV$, respectively. Here we show the results of AFM1 configuration for even SL, and the orbital magnetization of AFM2 simply flips in sign compared to AFM1. In Fig. \ref{fig:figS5} ($\mu=0eV$), when two surfaces are isolated ($t=0eV$), for even SL, the trivial part vanishes and the topological part increases linearly with electric potential $V_0$, and the slope represents a quantized magnetoelectric coefficient $\alpha=e^2/2h$. On the other hand, in odd SL both components of orbital magnetic moment are constant and the total magnetic moment is zero, consistent with the analytical results as shown previously. When there is coupling between two surfaces (Fig. \ref{fig:figS5}(b) and (d)), $\alpha$ remains zero for odd SL, while in even SL, the trivial orbital magnetization now presents a nonzero slope with an opposite sign compared to the topological part and thus the total $\alpha$ value is smaller than the quantized value $e^2/2h$, indicating the finite size effect in thin samples where the two surfaces are hybridized. In Fig. \ref{fig:figS6} ($\mu=0.1eV$), the trivial and topological parts for odd SL no longer cancel out with each other due to the nonzero chemical potential $\mu$ and there is a finite total orbital magnetic moment, while for even SL, the magnetic moment behavior is qualitatively similar to that of zero chemical potential case. We also examine more quantitatively the chemical potential $\mu$ dependence of the orbital magnetic moment in Fig. \ref{fig:figS7} at $V_0=10meV$. For even SL, both the trivial part and the topological part are constant in the gap, and when the chemical potential $\mu$ is in the conduction bands, there is a rapid change in the orbital magnetic moment due to the contribution from the conduction bands. For odd SL, the topological part varies linearly with $\mu$ while the trivial part is constant until the chemical potential reaches the conduction bands, which is consistent with the analytical results in Eq. (\ref{eq:mtrivial2}) and (\ref{eq:mtopo2}).

The orbital magnetic moments in the thin film model are also calculated, as discussed in the main text. The results are consistent with the two-surface-state model, that $\alpha=0$ for odd SL and $\alpha$ approaches quantized value $e^2/2h$ in even SL as the sample thickness increases, implying that the two surfaces become less coupled.

An essential difference between the magnetoelectric coefficient $\alpha$ and the axion parameter $\theta$ is that $\alpha$ is a physical quantity that can be measured in experiments and can take any values, while $\theta$ is a bulk topological quantity that will take two possible values 0 and $\pi$ mod $2\pi$ ($-\pi$ is equivalent to $\pi$) in the presence of time reversal symmetry (or other crystal symmetry such as inversion). Our calculation (discussed below) suggests that $\alpha$ mainly comes from surface state contribution and thus strongly depends on the properties of surfaces, and the difference between even and odd SLs is that the magnetization at the top and bottom surfaces is parallel or anti-parallel. In this sense, the even-odd effect of magnetoelectric coefficient $\alpha$ will still survives even when the layer number becomes very large, while $\theta$ is a bulk quantity defined in the infinite thickness limit and thus should be a constant value. 

\begin{figure}
    \centering
    \includegraphics[width=\textwidth]{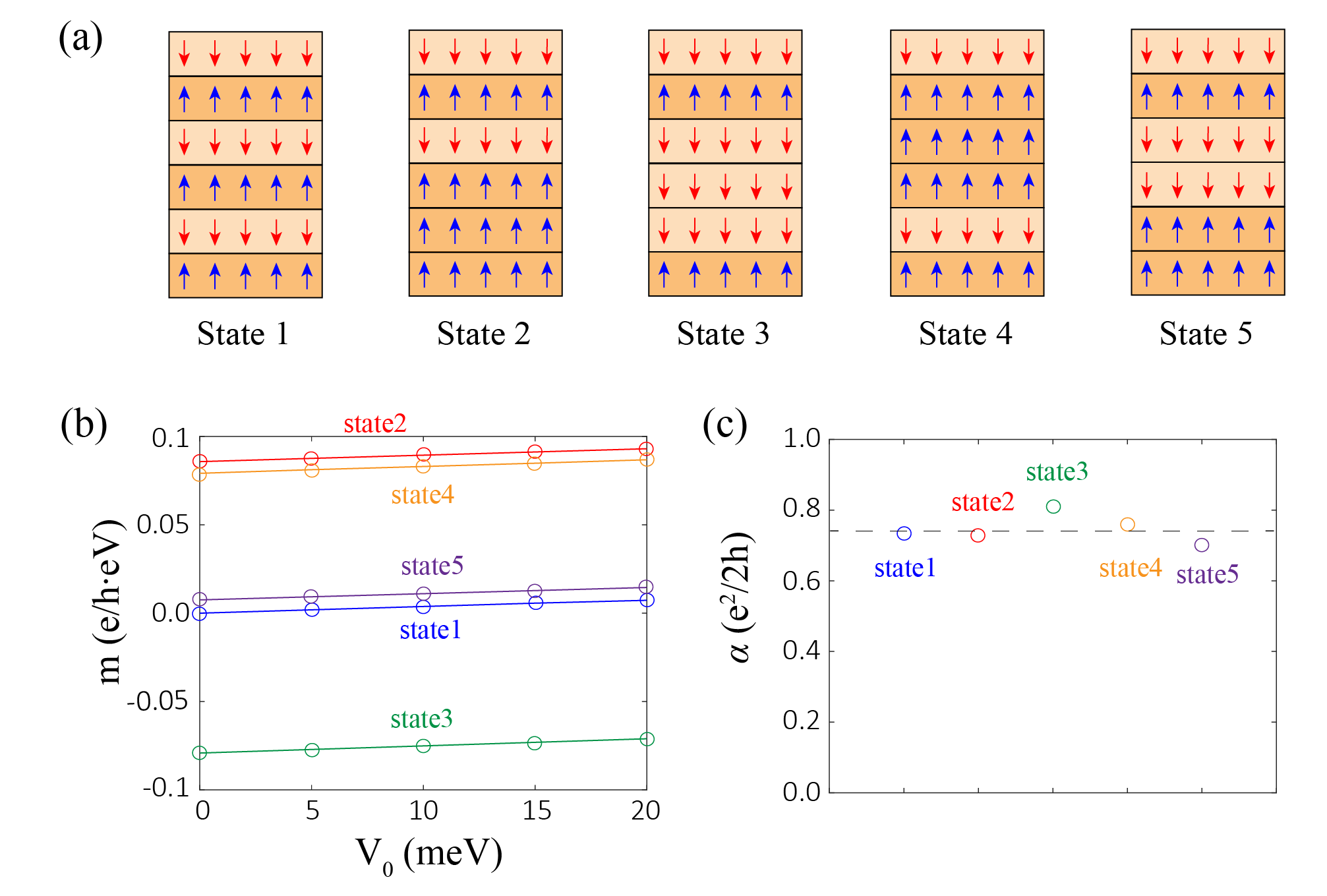}
    \caption{(a) Different magnetic states with the same surface magnetization and different bulk magnetization. (b) Calculated orbital magnetic moment $m$ as a function of $V_0$ for each state with $g=0.05eV$ and $\mu=0eV$. (c) Extracted magnetoelectric coefficient $\alpha$ for each state.}
    \label{fig:figS10}
\end{figure}

Our method is to directly evaluate the electron orbital magnetization $m$ of the thin films and then extract magnetoelectric coefficient $\alpha$ from the slope of orbital magnetization with varying the asymmetric potential $V_0$. To examine the influence of the surface and bulk layers numerically, we perform the following calculations for the magnetoelectric coefficient $\alpha$ in 6 SL MBT films with different magnetic configurations, where the surface magnetization remains the same as the AFM configuration while the bulk magnetization varies (Fig. \ref{fig:figS10}a). We can see that in Fig. \ref{fig:figS10}b for different states, the values of the total electron orbital magnetization are quite diverse, strongly depending on the bulk magnetic configurations. However, the extracted magnetoelectric coefficient $\alpha$ (which corresponds to the slope of the electron orbital magnetization $m$ respect to the asymmetric potential $V_0$) almost remains around the same value (Fig. \ref{fig:figS10}c), indicating that $\alpha$ mainly depends on the surface magnetization, regardless of the bulk magnetization properties.

In summary, although the overall orbital magnetization is determined by the bulk contribution, the slope of the orbital magnetization with respect to the asymmetric potential $V_0$ is mainly originated from the surface magnetization. Thus, the magnetoelectric coefficient $\alpha$ mainly comes from the surface state contribution, which is a combined effect from both the bulk axion parameter $\theta$ and the magnetizations at two surfaces. 

\end{document}